\documentclass[%
 reprint,
 amsmath,amssymb,
 aps,
]{revtex4-2}
\usepackage{booktabs}
\usepackage{threeparttable}
\usepackage{caption}
\usepackage{subcaption}
\usepackage{graphicx}
\usepackage{dcolumn}
\usepackage{bm}

\usepackage{multirow}
\begin{document}

\preprint{APS/123-QED}

\title{Effects of Dynamical Decoupling and Pulse-level Optimizations on IBM Quantum Computers}

\author{Siyuan Niu}
\email{siyuan.niu@lirmm.fr}
\author{Aida Todri-Sanial}%
 \email{aida.todri@lirmm.fr}
\affiliation{%
 LIRMM, University of Montpellier, France
}%

\date{\today}

\begin{abstract}
Currently available quantum computers are prone to errors. Circuit optimization and error mitigation methods are needed to design quantum circuits to achieve better fidelity when executed on NISQ hardware. Dynamical decoupling (DD) is generally used to suppress the decoherence error and different DD strategies have been proposed. Moreover, the circuit fidelity can be improved by pulse-level optimization, such as creating hardware-native pulse-efficient gates. This paper implements all the popular DD sequences and evaluates their performances on IBM quantum chips with different characteristics for various well-known quantum applications. Also, we investigate combining DD with pulse-level optimization method and apply them to QAOA to solve Max-Cut problem. Based on the experimental results, we found that DD can be a benefit for only certain types of quantum algorithms, while the combination of DD and pulse-level optimization methods always has a positive impact. Finally, we provide several guidelines for users to learn how to use these noise mitigation methods to build circuits for quantum applications with high fidelity on IBM quantum computers. 
	
\end{abstract}

\maketitle


\section{Introduction}

\label{sec:intro}
Quantum computing is rapidly growing in recent years and various technologies have been developed for different quantum platforms, the leading candidates being superconducting and trapped-ion devices. Several companies such as IBM, Rigetti, and IonQ provide publicly available cloud-based services which allow users to access their platforms remotely. But, today's quantum computers are still prone to errors caused by either unavoidable interactions with the environment or imperfect quantum controls. They are qualified as Noisy Intermediate-Scale Quantum (NISQ) computers~\cite{preskill2018quantum}. The largest quantum chip to date has 127 qubits released by IBM~\cite{ibmblog}. Quantum error correction (QEC) has been proposed to eliminate the noise impact and help achieve a fault-tolerant quantum device~\cite{calderbank1998quantum, terhal2015quantum}. However, the implementation of quantum error correction codes requires a large number of ancilla qubits, which is not feasible on current hardware. Therefore, alternative approaches are needed to tackle the noise issue. 

In order to reduce the noise impact and improve the quantum circuit fidelity, several quantum software were designed and they have made a great contribution to different circuit design processes, such as circuit synthesis~\cite{di2016parallelizing,zhang2004optimal}, or qubit mapping~\cite{niu2020hardware,li2019tackling}. In addition, quantum error mitigation (QEM) was introduced for error suppression on NISQ devices~\cite{endo2018practical}. There are different QEM techniques, such as readout error mitigation~\cite{bravyi2021mitigating}, dynamical decoupling (DD)~\cite{souza2012robust}, crosstalk mitigation~\cite{murali2020software,9516713}, zero-noise extrapolation (ZNE)~\cite{li2017efficient}, etc. Most methods require supplementary circuit executions to build the error map for mitigation. Whereas DD is one of the simplest strategies, which aims at mitigating the decoherence error (also called idle error) without any circuit overhead, and is the focus of this paper. The thrust of DD is to insert periodically a series of pulses to the idle qubits and return the qubits to their original states. There is a high probability to have idle qubits during execution due to the variation of gate latencies and limited parallelism caused by the anti-commutative gates. It has been shown in~\cite{das2021adapt} that the idle qubits can be almost 10 times more subject to errors if adjacent two-qubit operations are executing at the same time on IBM superconducting device. DD plays an important role in reducing the idle error and has been used in quantum volume experiments~\cite{jurcevic2021demonstration}, noise spectrum characterization~\cite{sung2019non}, and decoherence-protected quantum gate implementation~\cite{van2012decoherence}, etc. Moreover, there are different DD strategies, such as Hahn echo~\cite{solomon1958multiple}, CPMG~\cite{meiboom1958modified}, XY4~\cite{viola1999dynamical}, robust KDD~\cite{souza2011robust}, etc. They have diverse impacts on decoherence error suppression for different quantum devices. It has been shown in~\cite{pokharel2018demonstration} that DD is able to extend the lifetime of one-qubit states as well as entangled two-qubit states for IBM and Rigetti devices using XY4 sequence. But this sequence is shown to be vulnerable to experimental imperfections, while the use of robust DD was demonstrated to be capable of correcting the pulse errors on Rigetti devices~\cite{souza2021process}. 

Recently, IBM released Qiskit Pulse~\cite{mckay2018qiskit, alexander2020qiskit} allowing users to design and customize the gate pulse implementations. Some works have already attempted to  optimize the pulse controls and reduce the pulse durations with the help of Qiskit Pulse~\cite{shi2019optimized, khaneja2005optimal, egger2014adaptive}. However, such method often needs additional calibrations which is time-consuming and requires a deep familiarity with quantum control. In~\cite{stenger2021simulating, earnest2021pulse}, the authors proposed a new technique to create more hardware-native pulse-efficient gates, which improves the gate fidelity without the overhead of extra calibrations.

So far, the aforementioned techniques were only tested separately for limited benchmarks and quantum hardware. Several questions still remain unclear: (1) What are the impacts of different DD sequences on specific quantum algorithms? (2) For a certain benchmark, does the impact of different DD sequences vary across different quantum chips? (3) Will the combination of DD and pulse-level optimization methods further improve the circuit fidelity? In our work, we address these questions and our main contributions can be listed as follows:

\begin{itemize}
	\item We explore the performance of different DD sequences on various quantum applications and evaluate the experiments on several IBM devices with different qubit numbers and quantum volumes. To the best of our knowledge, this is the first attempt to illustrate the behavior of applying robust KDD sequence to IBM quantum devices.
	
	\item We combine the DD technique with pulse-efficient optimization method to demonstrate their benefits on Quantum Approximate Optimization Algorithm (QAOA) to solve Max-Cut problem. 
	
	\item Based on the experimental results, we provide guidelines and insights for users to apply application-oriented dynamical decoupling and pulse-level optimization techniques.
	
\end{itemize}

\section{Background}

\subsection{Dynamical decoupling}
Dynamical decoupling is widely used in suppressing the decoherence error by reducing the interaction between the system and the environment.  Considering a system-bath Hamiltonian $H$ shown in~\eqref{eq:1}, $H_s$ and $H_B$ being the Hamiltonian of the system and the bath respectively, and the interacting term being $H_{SB}$ (see~\eqref{eq:2}), where $\sigma_{i}^{\alpha}$ is the Pauli matrix acting on qubit $i$, $B_{i}^{\alpha}$ is the operator of the environment, and $\alpha \in \{x, y, z\}$. DD aims at reducing the impact of system-environment interaction and various DD protocols have been developed to improve the performance of quantum computers. Here, we review the main strategies of DD implementations and a summary of these DD sequences is shown in Table~\ref{tab:1}.

\begin{equation}
	H = H_s + H_B + H_{SB}
	\label{eq:1}
\end{equation}

\begin{equation}
	H_{SB} = \sum_{i=1}^{N} \sum_{\alpha \in \{ x,y,z \} } \sigma_{i}^{\alpha} \otimes B_{i}^{\alpha}
	\label{eq:2}
\end{equation}

\begin{table}[!htp]
	\caption{\label{tab:dd}A summary of different DD sequences. }
	\centering
	\resizebox{\linewidth}{!}{%
	\begin{threeparttable}
		\begin{tabular}{c c }
			\toprule                              
			DD sequences & Pulse implementation\\
			\midrule
			Hahn echo~\cite{solomon1958multiple} & $X$ or $Y$\\
			\midrule
			CP/CPMG~\cite{carr1954effects,meiboom1958modified} & $(X)^n$ or $(Y)^n$ \\
			\midrule
			XY4~\cite{viola1999dynamical} & $(X-Y-X-Y)^n$\\
			\midrule
			XY8~\cite{gullion1990new} & $(XY4-\overline{XY4})^n$\\
			\midrule
			XY16~\cite{gullion1990new} & $(XY8-\overline{XY8})^n$\\
			\midrule
			\multirow{2}{*}{UDD~\cite{uhrig2007keeping}} & $(X)^n$ or $(Y)^n$ with space $\tau$ \\
			& where $\tau = \sin^{2} (\frac{\pi j}{2n+2}), j \in \{1, 2, ...n\}$\\
			\midrule
			\multirow{3}{*}{KDD~\cite{souza2011robust}} & $(\pi)_{\frac{\pi}{6}+\phi}-(\pi)_{\phi}-(\pi)_{\frac{\pi}{2}+\phi}-(\pi)_{\phi} - (\pi)_{\frac{\pi}{6}+\phi}$ \\
			& where $(\theta)_{\phi} = e^{\frac{i \theta}{2}[\cos(\theta)\sigma_x + \sin(\theta)\sigma_y]}$, \\
			& $\phi = 0, \frac{\pi}{2}, 0, \frac{\pi}{2}$ \\
			\bottomrule
		\end{tabular}
		\begin{tablenotes}
			\item $n$: repetition time of the basic DD cycle.
			\item $(\theta)_{\phi}$: a rotation of $\theta$ around the axis defined by $\phi$.
		\end{tablenotes}
	\end{threeparttable}
}
	\label{tab:1}
\end{table}

\begin{itemize}
	\item \textbf{Hahn echo.} The spin-echo sequence~\cite{solomon1958multiple} is used to reduce the inhomogeneous effects from the environmental magnetic field. It applies a $\pi$ pulse to the spin system to inverse the spins after a period of time $t$ and let the system be refocused during the same duration. At that time, $T_2^{*}$ effects are removed and we can obtain a $T_2$ echo.
	
	\item \textbf{CP and CPMG.}
	Carr and Purcell proposed a series of $\pi$ pulses separated by a constant interval, known as the CP sequence, to further reduce the effect of self-diffusion in the inhomogeneous field~\cite{carr1954effects}.    
	But, extra pulses can introduce more errors and destroy the state of the system. Therefore, Meiboom and Gill improved the CP pulse by developing the CPMG sequence~\cite{meiboom1958modified}, retaining the CP pulse but introducing an additional phase shift in the first pulse to compensate the pulse errors.     
	%
	\item \textbf{UDD.} Hahn echo and CP/CPMG are equidistant pulses. While UDD aims at optimizing the $\pi$ sequence based on CP/CPMG by varying the intervals between each pulse~\cite{uhrig2007keeping}. It is proven to be the optimal pulse to suppress low frequency noise and is insensitive to thermal fluctuations. It outperforms the equidistant DD sequences especially for systems whose spectral densities have high frequencies with sharp cutoff. For other more general cases, equidistant DD sequences were shown to perform better~\cite{ajoy2011optimal}.

	\item \textbf{XY4.} All the DD sequences above only rotate around one single axis. They are exclusively useful when the system-environment interaction is orthogonal to the rotation axis. XY4 is the simplest DD sequence to generally suppress system-environment interaction along three directions~\cite{viola1999dynamical}. It inserts alternatively $\pi$ rotations around X and Y axes independently of the initial state.

	\item \textbf{XY8 and XY16.} 
	DD can sometimes detriment the fidelity due to the accumulation of errors caused by pulse imperfections. One approach to reducing the errors is to combine the basic sequence with its inverse for self-correction so that the unwanted terms can be canceled~\cite{gullion1990new}. XY4 is usually chosen as the basic cycle. 
	XY8 is composed of XY4 sequence and its inverse while XY16 contains XY8 and its inverse.

	\item \textbf{KDD.} The other approach to avoid pulse imperfections is to replace each pulse of a DD sequence by a robust composite pulse, which is designed to generate ideal rotations even if there exist pulse imperfections~\cite{grant1996encyclopedia}. KDD constructs a DD block using a 5-pulse composite $\pi$ pulse, and combine two of such DD block with and without a phase shift $\pi/2$ as $(KDD_{\phi}, KDD_{\phi+\pi/2})^2$~\cite{souza2011robust}. It is composed of 20 pulses in total. 
	
\end{itemize}

DD has been demonstrated to have the capability to mitigate the decoherence errors on IBM and Rigetti platforms using XY4 sequences~\cite{pokharel2018demonstration}. The experiment is performed as follow: it prepares different initial states by varying the angles of the rotation gates, inserts XY4 sequences, and compares the difference between input and output states. The impacts of different DD sequences on Rigetti device are reported in~\cite{souza2021process}. A large number of DD sequences are evaluated, including XY4, XY8, KDD, etc., and quantum process tomography is used to characterize the evolution. KDD is shown to be the most robust pulse sequence against pulse imperfections.  Both papers evaluate DD sequences on simple benchmarks instead of real quantum applications, and there is no experiment to illustrate applying KDD to IBM quantum devices. Moreover, the experiments in~\cite{das2021adapt} show that the naive implementation of DD (inserting DD sequences to all the idle qubits when it is possible) can not always improve the circuit fidelity. Therefore, Das et al. proposed an Adaptive Dynamical Decoupling framework to estimate the DD impact for each circuit and adjust DD sequence to ensure it improves the circuit fidelity~\cite{das2021adapt}. This method achieves fidelity improvement but introduces a large overhead of DD impact characterization for a given application. Also, Ravi et al. proposed VAQEM~\cite{ravi2021vaqem}, an approach that dynamically inserts DD sequences for variational algorithm with the overhead of tuning DD sequence. In our paper, instead of carefully adjusting DD sequences for better error mitigation with extra circuit overhead, we exhaustively evaluate DD sequences on extensively used quantum applications, to give a high-level idea about if general DD techniques can really be beneficial for these applications. It is also the first attempt to  implement KDD on IBM devices to check its performance. The experiments are performed on various IBM devices with different qubit numbers and quantum volumes to provide general insights about application-oriented DD noise mitigation on IBM quantum devices.

\subsection{Pulse-efficient technique}
$CNOT$ gate is the only two-qubit operation included in the basis gates for IBM quantum devices and its calibrations are provided through the IBM quantum dashboard. It is implemented by a $R_{ZX}(\frac{\pi}{2})$ gate along with some single-qubit gates~\cite{chow2014implementing} on the device.
$R_{ZX}$ gate is realized by the echoed cross-resonance gate~\cite{sheldon2016procedure}, which is specific for the IBM fixed-frequency superconducting transmon qubit device due to its low overhead and high fidelity. When executing a circuit on IBM quantum hardware, every multi-qubit operation needs to be transpiled to the $CNOT$ basis which is not flexible and less efficient. 

Therefore,~\cite{stenger2021simulating, earnest2021pulse} proposed a pulse-efficient circuit transpilation framework and has shown to achieve higher fidelity than $CNOT$ basis transpilation for certain benchmarks. Instead of only using a $R_{ZX}(\theta)$ with a fixed angle $\theta = \frac{\pi}{2}$ for $CNOT$ gate, a flexible echoed $R_{ZX}$ gate is implemented with arbitrary angle and $R_{ZX}(\theta) = X R_{ZX}(-\frac{\theta}{2}) X R_{ZX}(\frac{\theta}{2})$ to enable transpilations to $R_{ZX}$ basis. This method does not require any additional pulse calibration since the calibration of $R_{ZX}$ gate can be easily calculated from the $CNOT$ calibrations. 

\begin{figure}[h]
	\centering
	\includegraphics[scale=0.6]{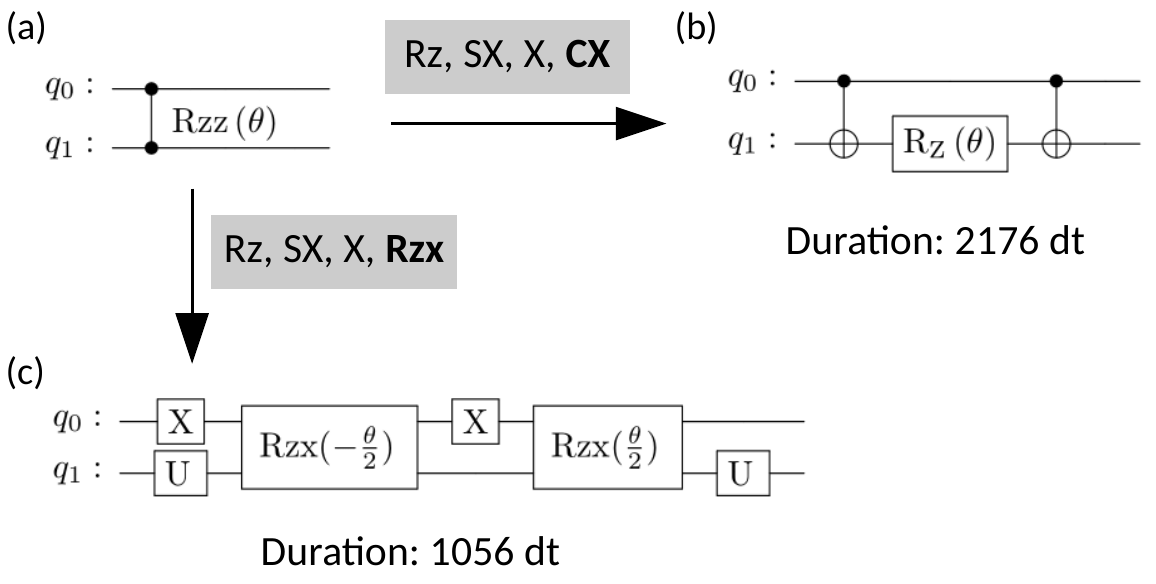}
	\caption{(a) Original $R_{ZZ}$ gate with $\theta = \frac{\pi}{3}$. The circuit is mapped to the first two qubits of IBM Q 27 Toronto. (b) The transpilation result with basis gate sets $\{R_Z, SX, X, CX\}$, and the duration is 2176 dt. dt is the sampling time for Qiskit pulses and is set to 0.2 ns. (c) The transpilation result with basis gate sets $\{R_Z, SX, X, R_{ZX}\}$, and the duration is 1025 dt. We replace $R_Z(\frac{m \pi}{2}) \sqrt{X} R_Z(\frac{n \pi}{2})$ by $U$ to simplify the circuit. $m, n \in \mathbb{R}$ and can be varied across different transpilation passes.}
	\label{fig:pe}
\end{figure}

An example of transpiling $R_{ZZ}$ gate to two different basis gate sets is shown in Fig.~\ref{fig:pe}. We first transpile the $R_{ZZ}$ gate to the IBM $CNOT$-basis gate sets (Fig.~\ref{fig:pe}(b)) and then transpile again the original circuit to $R_{ZX}$-basis gate sets (Fig.~\ref{fig:pe}(c)). 
The duration of the second circuit is reduced by 51.1\%  compared to the first gate sets. 
This technique exposes the echo of the cross-resonance gates which enables at most one single-qubit gate between each non-echoed $R_{ZX}$ gate so that it can shorten the total circuit duration. 


Quantum algorithms requiring a lot of two-qubit control-rotation gates, such as $R_{ZZ}$ or $R_{YY}$, which can be directly compiled to $R_{ZX}$ gates along with some single-qubit gates, are particularly benefited from this pulse-efficient transpilation framework, for example QAOA~\cite{farhi2014quantum}, quantum hamiltonian simulation~\cite{papageorgiou2012efficiency}, etc.

\begin{figure*}
	\centering
	\subfloat[Bernstein-Vazirani (BV) algorithm\label{fig:bv}]{\includegraphics[scale=0.9,width=0.3\textwidth]{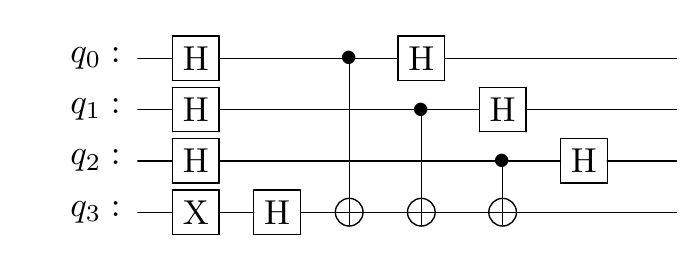}}
	\hfil
	\subfloat[Hidden Shift (HS) algorithm\label{fig:hs}]{\includegraphics[scale=0.9,width=0.3\textwidth]{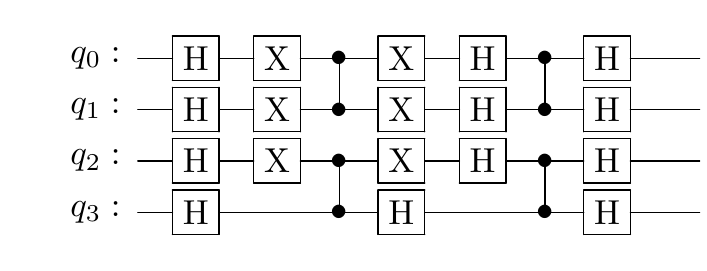}}
	\hfil
	\subfloat[Graph State (GS)\label{fig:gs}]{\includegraphics[scale=0.9,width=0.22\textwidth]{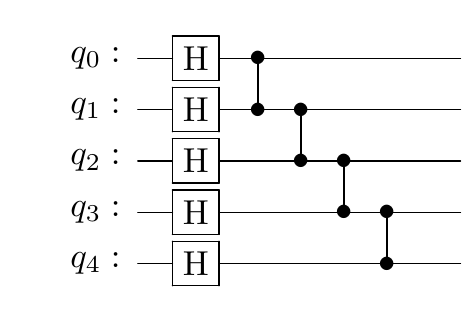}}
	
	\subfloat[Quantum Fourier Transform (QFT)\label{fig:qft}]{\includegraphics[scale=0.9]{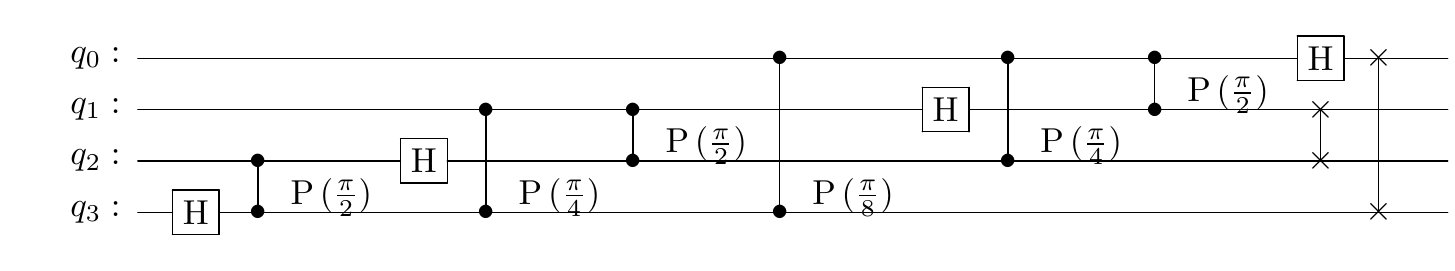}}
	
	\caption{Examples of different quantum circuits.
		Note that, the Graph State circuit is constructed linearly based on the target hardware topology. }
	\label{fig:DD}
\end{figure*}

\section{Methods}
We construct DD sequences according to Table~\ref{tab:dd}, and $n$ is the number of repetition time for the basic DD pulse. For Hahn echo, we insert one $X$ or $Y$ gate during the idle time. We set $n=2$ for CP/CPMG, CP (resp.CPMG) being implemented as the sequence $X-X$ (resp. $Y-Y$). For XY4, XY8, and XY16, $n$ is set to 1 so that the sequence of pulses is inserted once. All the aforementioned DD sequences are equidistant and symmetrical: there is a delay of  $\tau/2$ at the beginning and the end of the idle time, and a delay of $\tau$ between each DD pulse. Whereas UDD is composed of non-equidistant pulses, it acts differently from CPMG for all $n>2$. As there is no clear conclusion about which repetition number $n$ we should pick to have the best performance of UDD on IBM quantum devices, we set $n=8$ as demonstrated in Qiskit tutorial. The repetition gate is set to $X$ and $Y$, and is marked as UDD\_X and UDD\_Y in the results. For KDD, $(\theta)_{\phi}$ represents a rotation of $\theta$ around the axis defined by $\phi$. Since $\theta_{\phi} = R_Z(\phi)R_X(-\theta)R_Z(-\phi)$, we insert 60 gates in total, where delay is only introduced between each $\theta_{\phi}$, which means there is no space between the three gates constructing $\theta_{\phi}$. Note that, since $R_Z$ gate is free on IBM quantum devices, KDD is translated to ``only" 20 pulses in the end.

First, we evaluate the DD effects on IBM quantum computers by applying different DD sequences to various well-known quantum applications, including Bernstein-Vazirani (BV) algorithm, Hidden Shift (HS) algorithm, Quantum Fourier Transform (QFT), and Graph State (GS). The basis quantum circuit structures of these quantum applications are shown in Fig.~\ref{fig:DD}. BV algorithm implements an oracle function $f(x)$, which represents the dot product between $x$ and a secret string $s$, with the objective of finding $s$. HS algorithm constructs an oracle that encodes two functions $f$ and $g$, and there exists a secret string $s$ such that $g(x) = f(x+s)$. The goal of the algorithm is to find $s$. QFT is the quantum version of discrete Fourier transform, and is the essential part for many other quantum algorithms, such as Shor's algorithm~\cite{shor1999polynomial}, quantum phase estimation algorithm~\cite{kitaev1995quantum}, etc. Graph State is a quantum state prepared based on a graph. Specifically, we can build a Graph State according to a given hardware topology, where there is an edge on the graph when the two qubits are coupled, and the connection is represented by a $CZ$ gate. It can entangle all the qubits of the device and is important for error correction. 

\begin{figure}[h]
	\centering 
	\subfloat[3-reg-4\label{3-reg-4}]{\includegraphics[scale=0.3, width=0.35\linewidth]{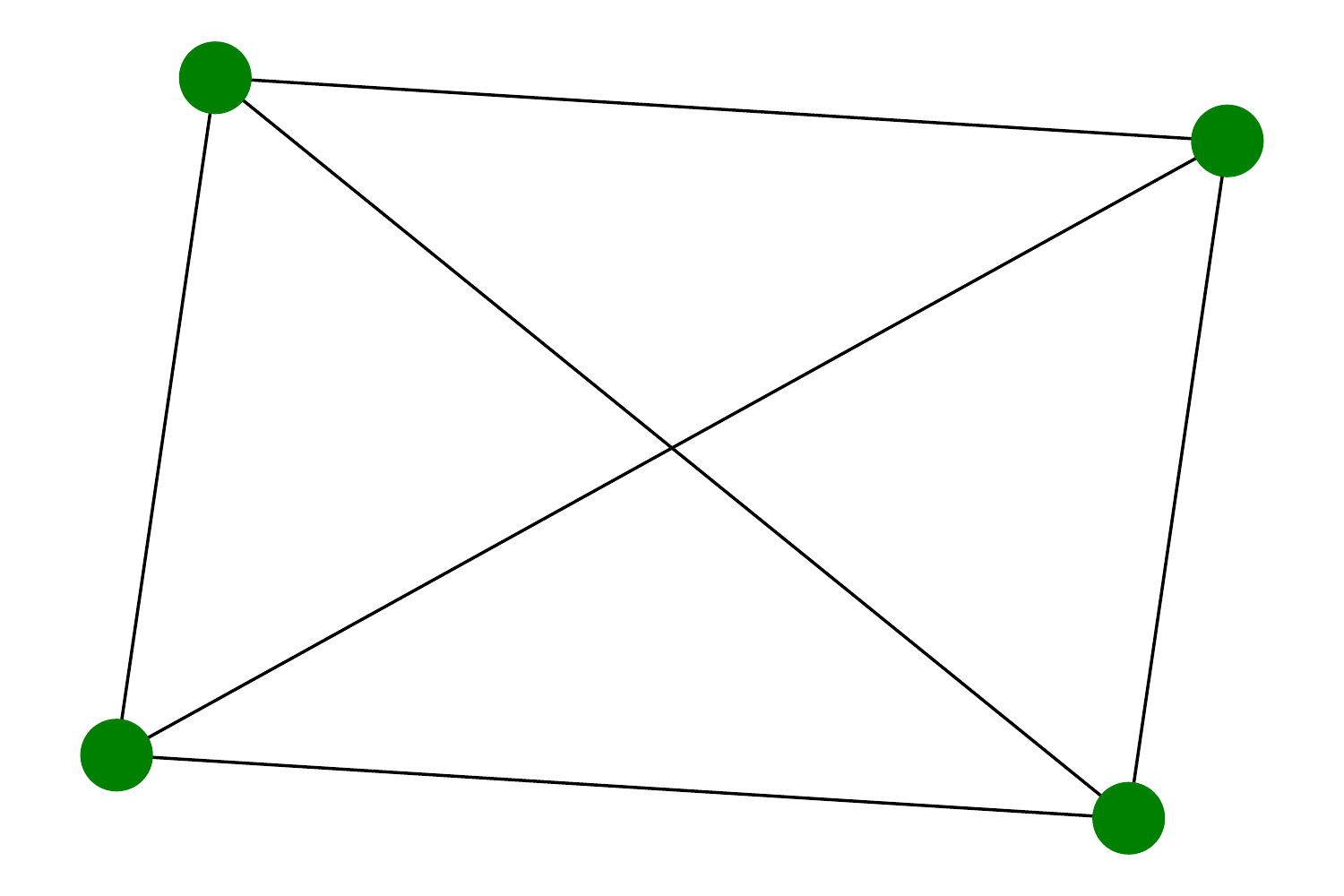}}
	\hspace{10mm}
	\subfloat[rand-4-0.5\label{rand-4-0.5}]{\includegraphics[scale=0.3,width=0.35\linewidth]{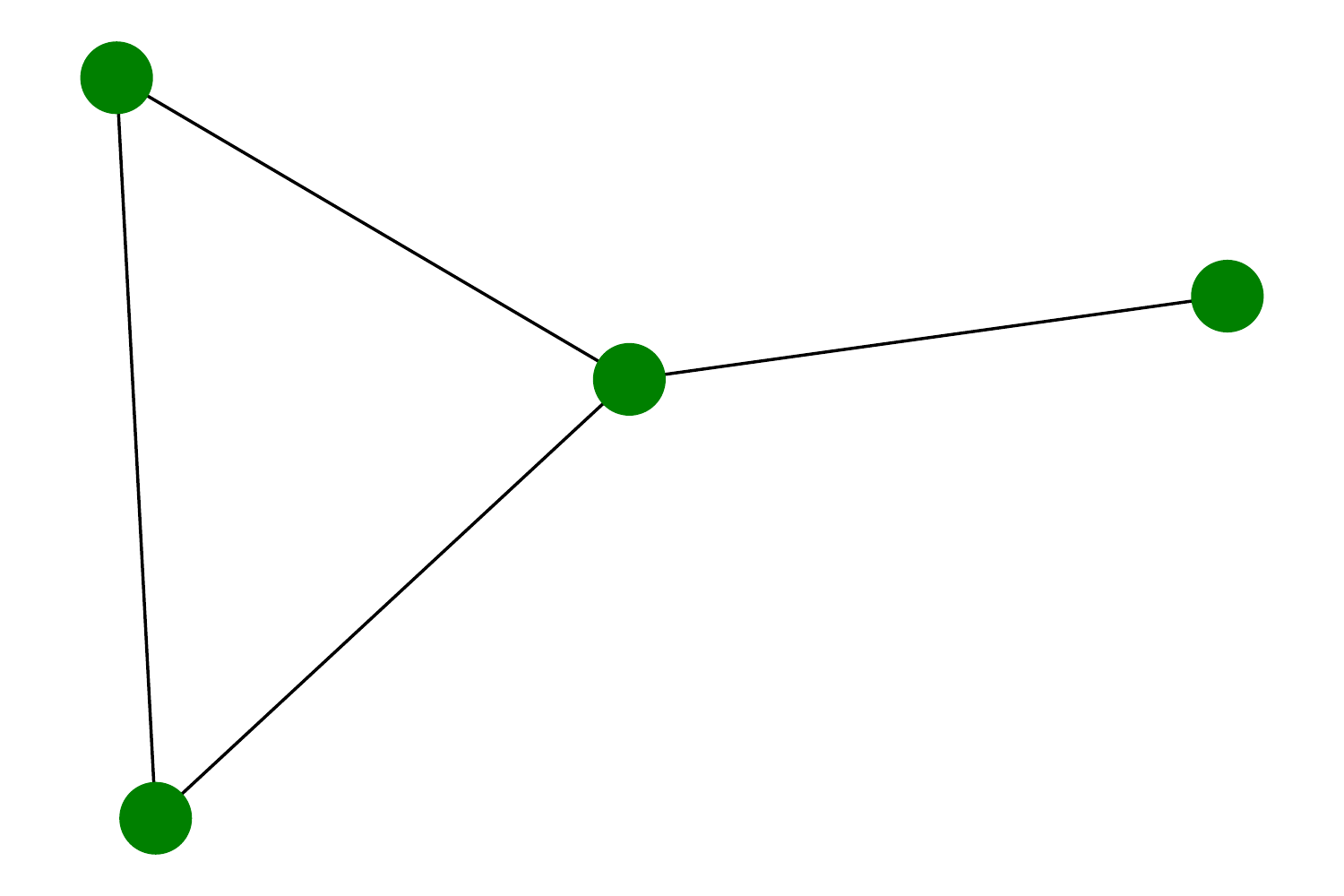}}
	
%
	\caption{An example of (a) 3-regular graph with degree of 4, (b) random graph with degree of 4 and edge creation probability of 0.5.}
	
	\label{fig:graphs}
\end{figure}

Second, we evaluate the performance of combining different DD sequences with pulse-efficient transpilation technique. We apply them to QAOA to solve the Max-Cut problem, since QAOA ansatz is composed of $R_{ZZ}$ gates along with some single-qubit gates, which can be profited from the pulse-efficient method.
We generate randomly 3-regular graphs using ReCirq~\cite{quantum_ai_team_and_collaborators_2020_4091470} and random graphs using Networkx with different degrees as our benchmarks. For example, we use \emph{3-reg-4} to represent a 3-regular graph with degree of 4 (qubits) and \emph{rand-4-0.5} for a random graph with 4 nodes/qubits and a probability for edge creation of 0.5 (see Fig.~\ref{fig:graphs}).

We choose the following metrics for different benchmarks to demonstrate the impact of DD and pulse-level optimization technique on application fidelity.
\begin{itemize}
	\item Probability of Successful Trial (PST)~\cite{tannu2019not}.
	This metric is defined by the ratio of the number of trials that give the expected result to the total number of trials, and higher is better. It is dedicated to the benchmarks with one certain correct result, such as BV algorithm and HS algorithm. 
	
	\item Jensen-Shannon Divergence (JSD). It is used to measure the similarity between two probability distributions, and lower is better. It is suitable for Graph State circuit and QFT whose output is a distribution.
	
	\item Approximation ratio~\cite{harrigan2021quantum}. It is specifically designed to evaluate the performance of QAOA circuit and is defined as
	$\langle C \rangle$/$C_{min}$, where $\langle C  \rangle$ is the expectation value obtained by the quantum computer and  $C_{min}$ is calculated by the classical \emph{NumPyMinimumEigensolver}. We aim at maximizing the approximation ratio and 1 means a perfect solution. 
\end{itemize}

\begin{figure*}[h]
	\centering 
	\subfloat[ibmq\_jakarta\label{fig:bv_jakarta}]{\includegraphics[scale=0.28,width=0.4\textwidth]{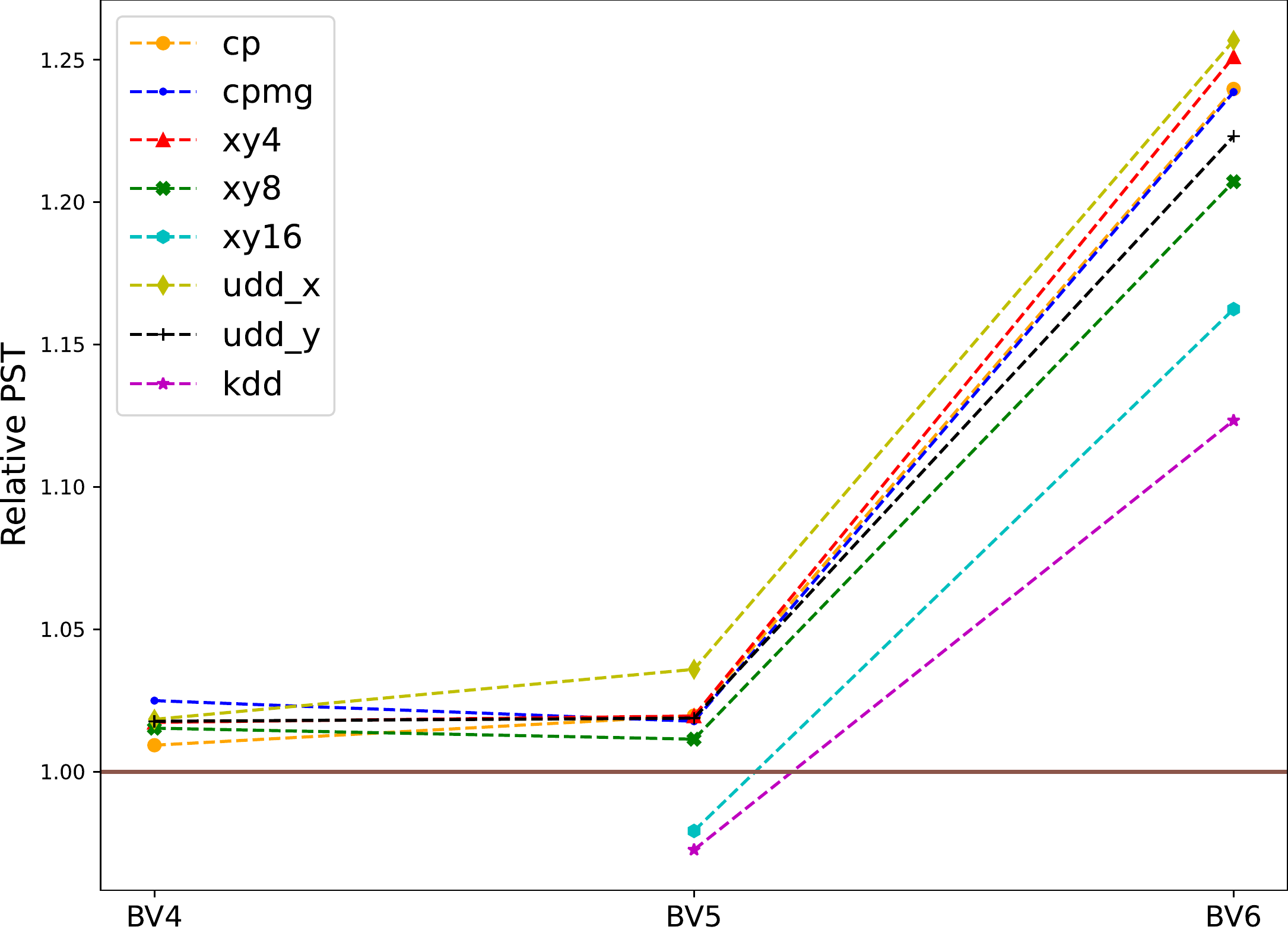}}
\hfil
    \subfloat[ibmq\_guadalupe\label{bv_guadalupe}]{\includegraphics[scale=0.28,width=0.4\textwidth]{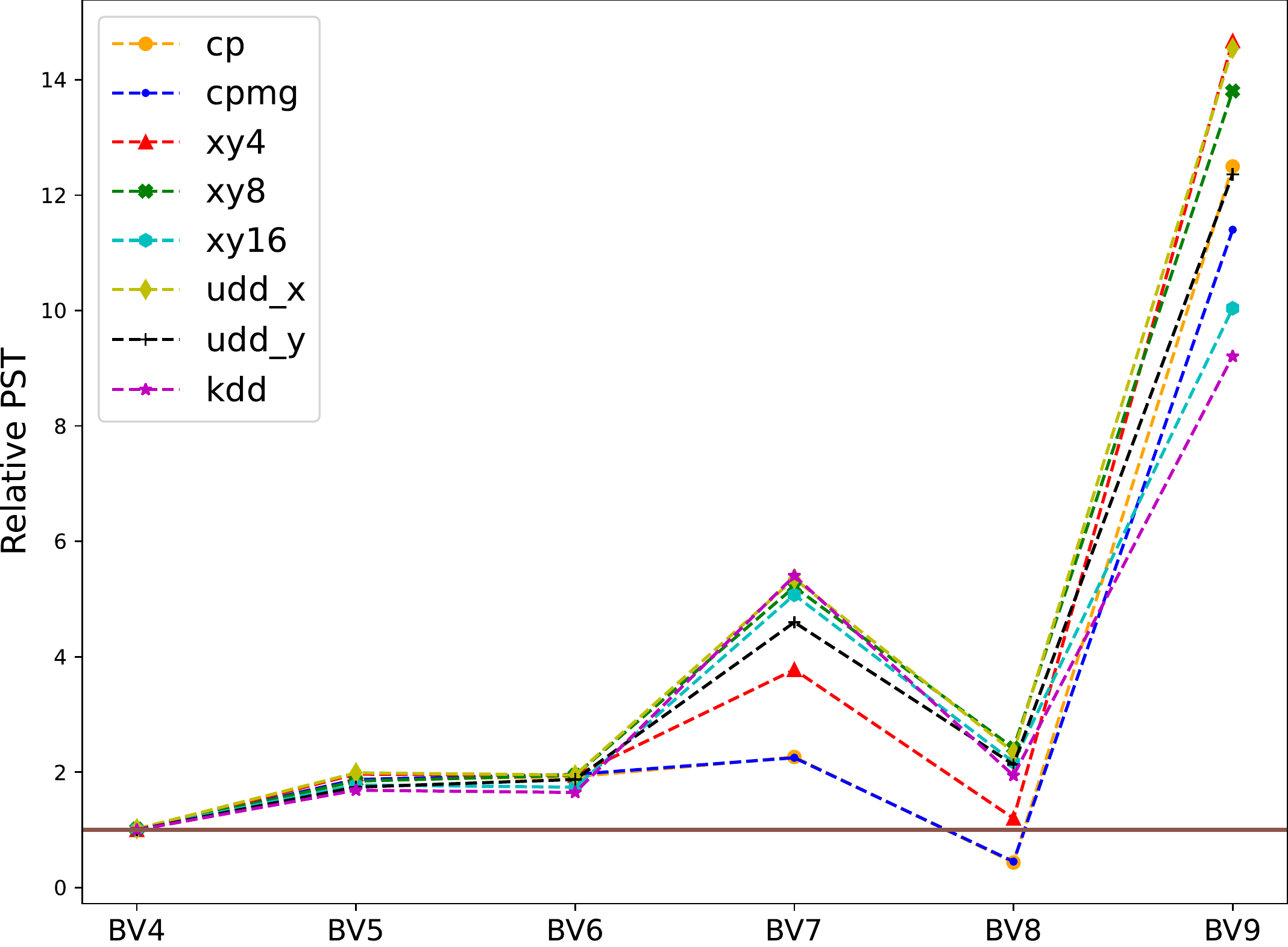}}
	
	\subfloat[ibmq\_toronto\label{bv_toronto}]{\includegraphics[scale=0.28,width=0.4\textwidth]{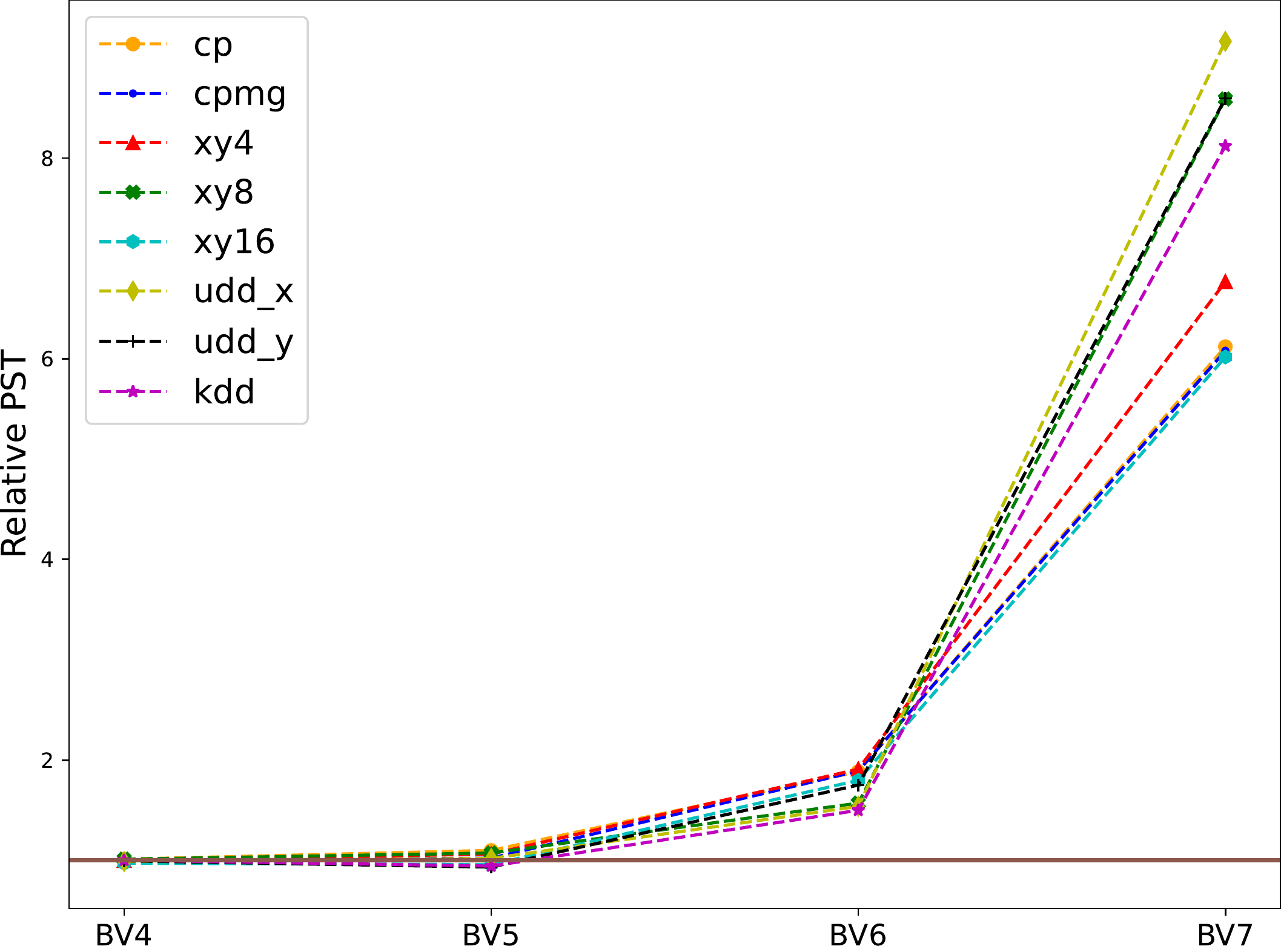}}
\hfil
	\subfloat[ibmq\_montreal\label{bv_montreal}]{\includegraphics[scale=0.28,width=0.4\textwidth]{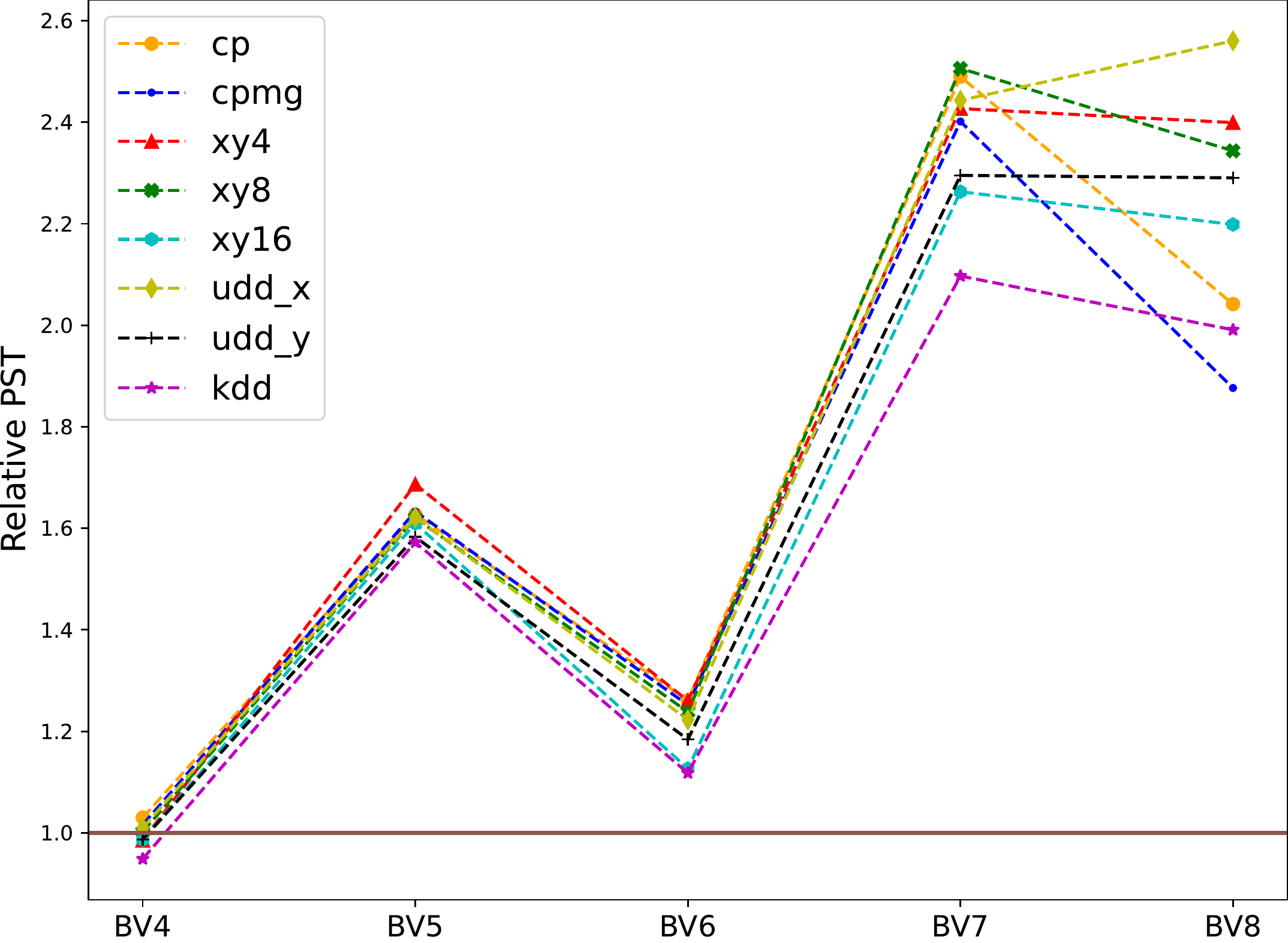}}

	\caption{Relative PST results for BV circuits on different IBM quantum devices. Higher is better.}
	\label{fig:bvs}
\end{figure*}

\begin{figure*}[h]
	\centering 
	\subfloat[ibmq\_jakarta	\label{qft_jakarta}]{\includegraphics[scale=0.28,width=0.4\textwidth]{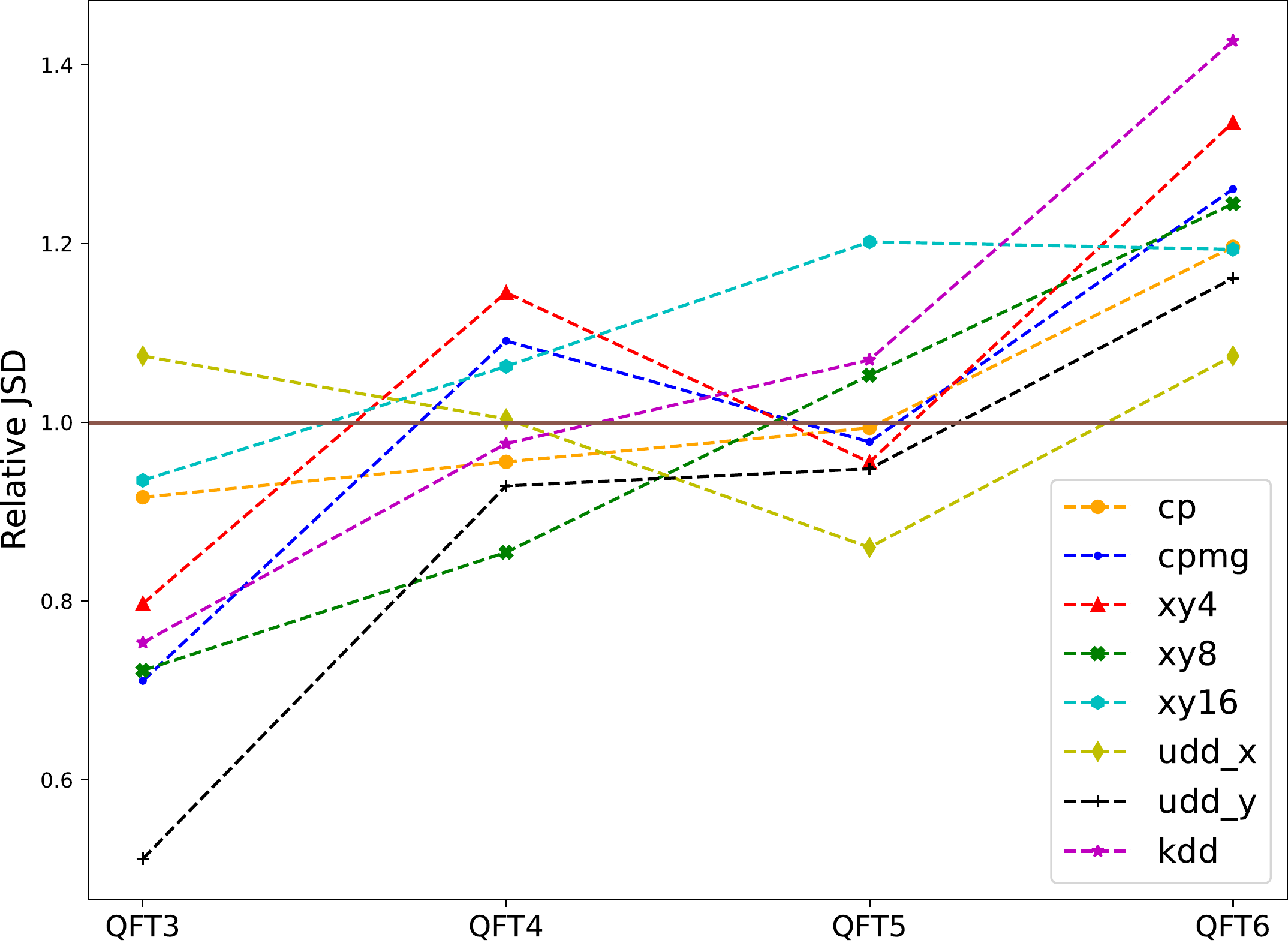}}
\hfil
	\subfloat[ibmq\_guadalupe\label{qft_guadalupe}]{\includegraphics[scale=0.28,width=0.4\textwidth]{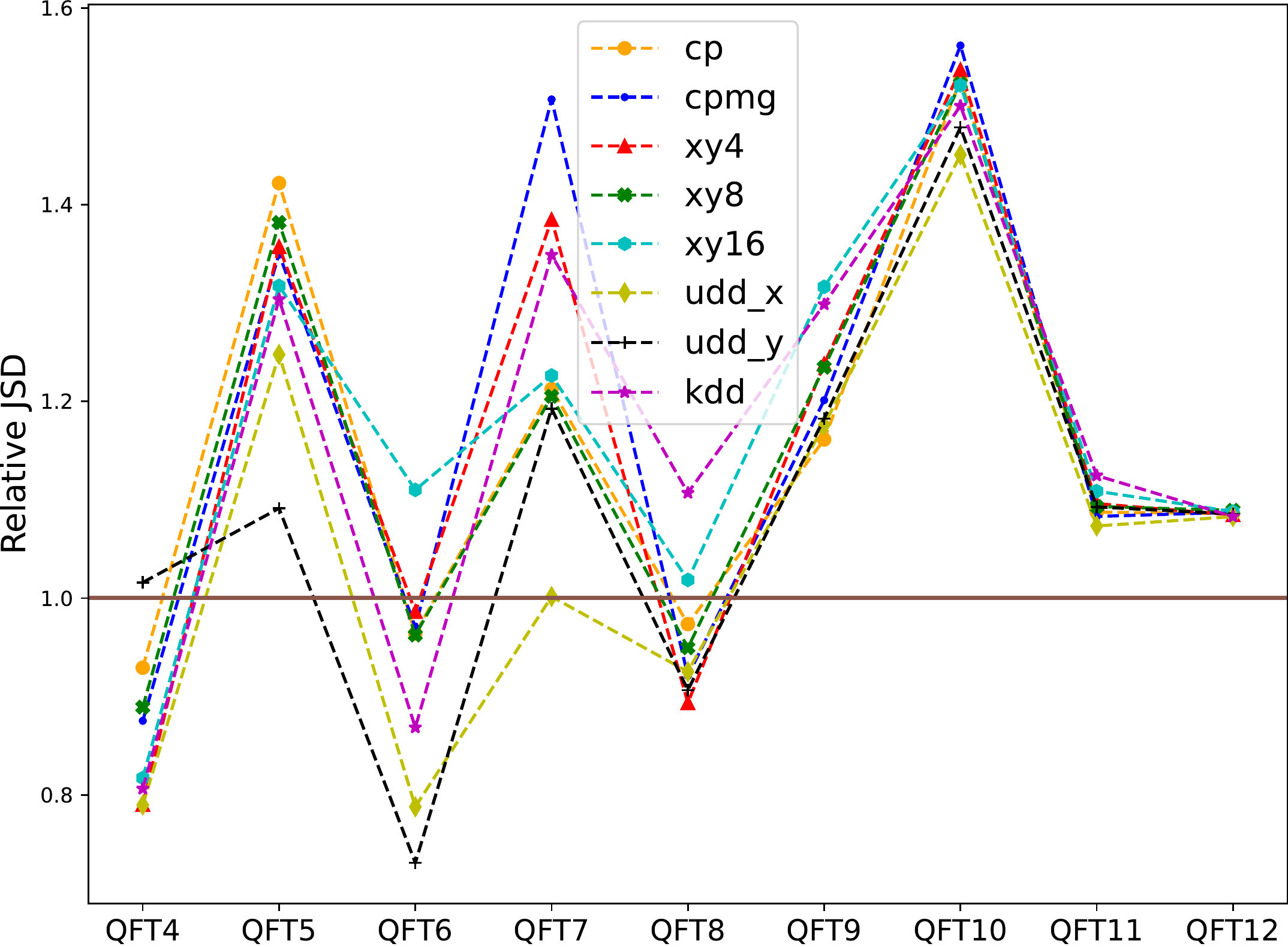}}

	\subfloat[ibmq\_toronto\label{qft_toronto}]{\includegraphics[scale=0.28,width=0.4\textwidth]{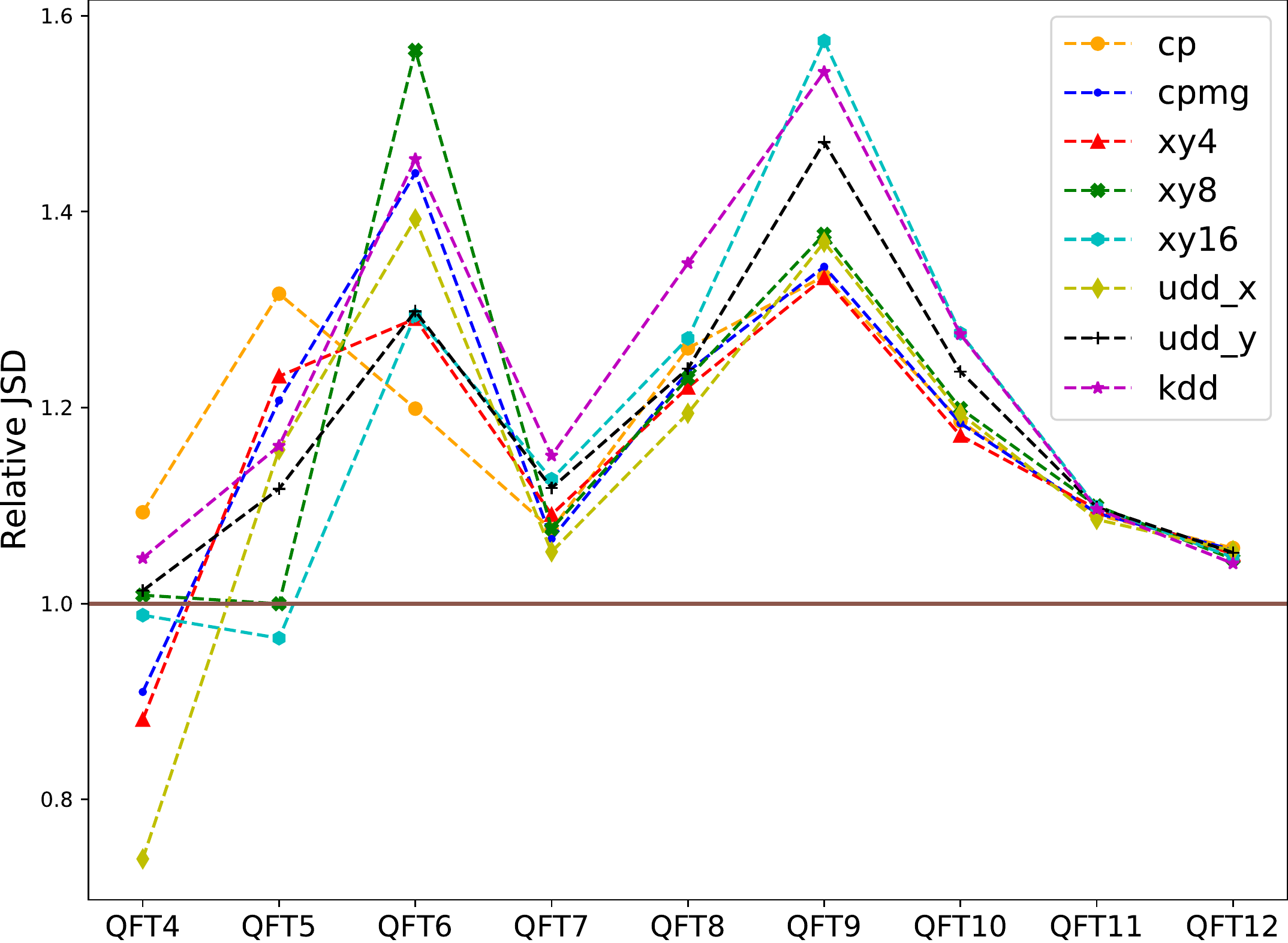}}
\hfil
	\subfloat[ibmq\_montreal\label{qft_montreal}]{\includegraphics[scale=0.28,width=0.4\textwidth]{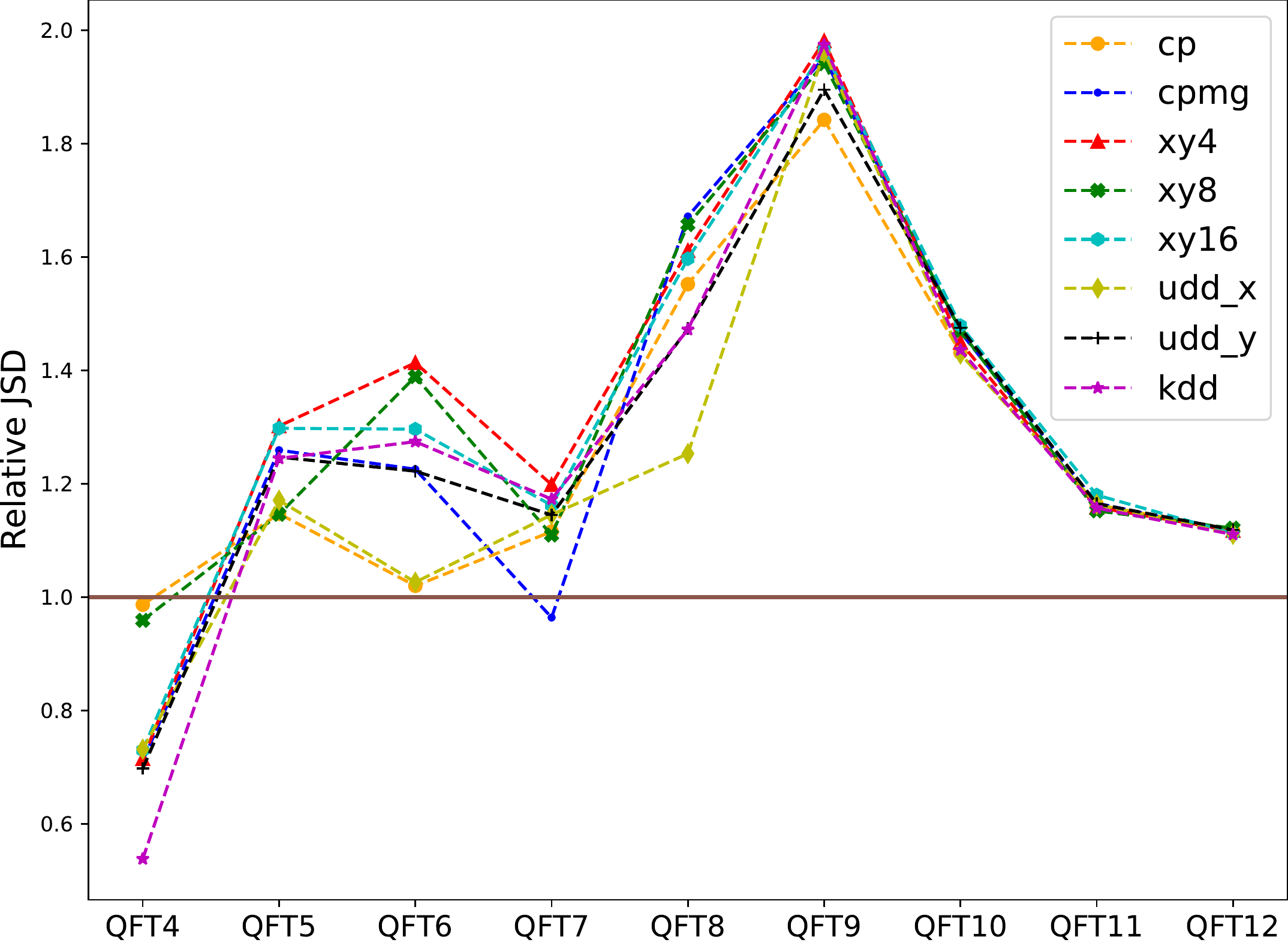}}

	\caption{Relative JSD results for QFT circuits on different IBM quantum devices. Higher is better.}
	\label{fig:qfts}
\end{figure*}

\begin{figure*}[h]
	\centering 
	\subfloat[ibmq\_jakarta\label{gs_jakarta}]{	\includegraphics[scale=0.28,width=0.4\textwidth]{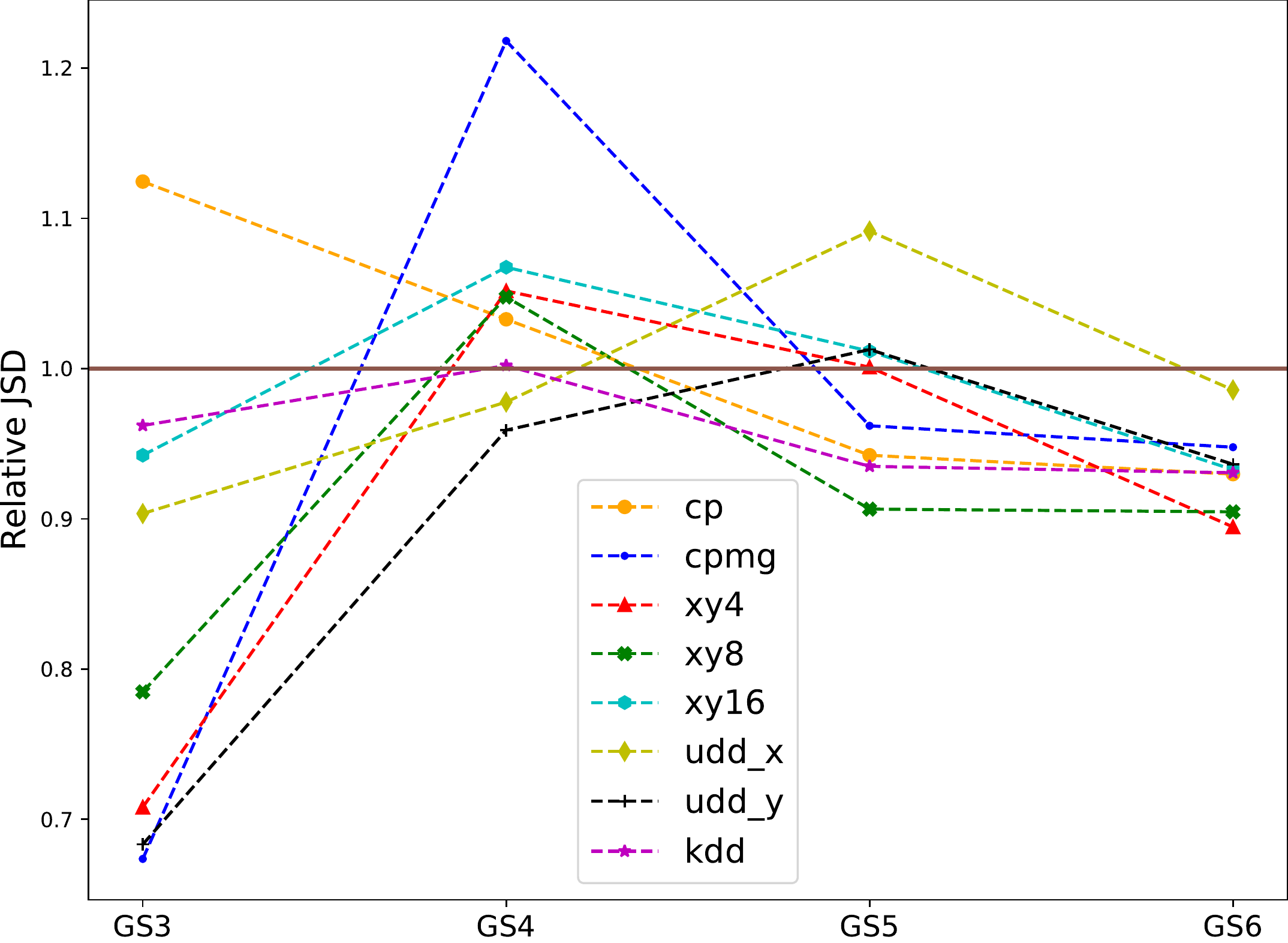}}
\hfil
	\subfloat[ibmq\_guadalupe\label{gs_guadalupe}]{\includegraphics[scale=0.28,width=0.4\textwidth]{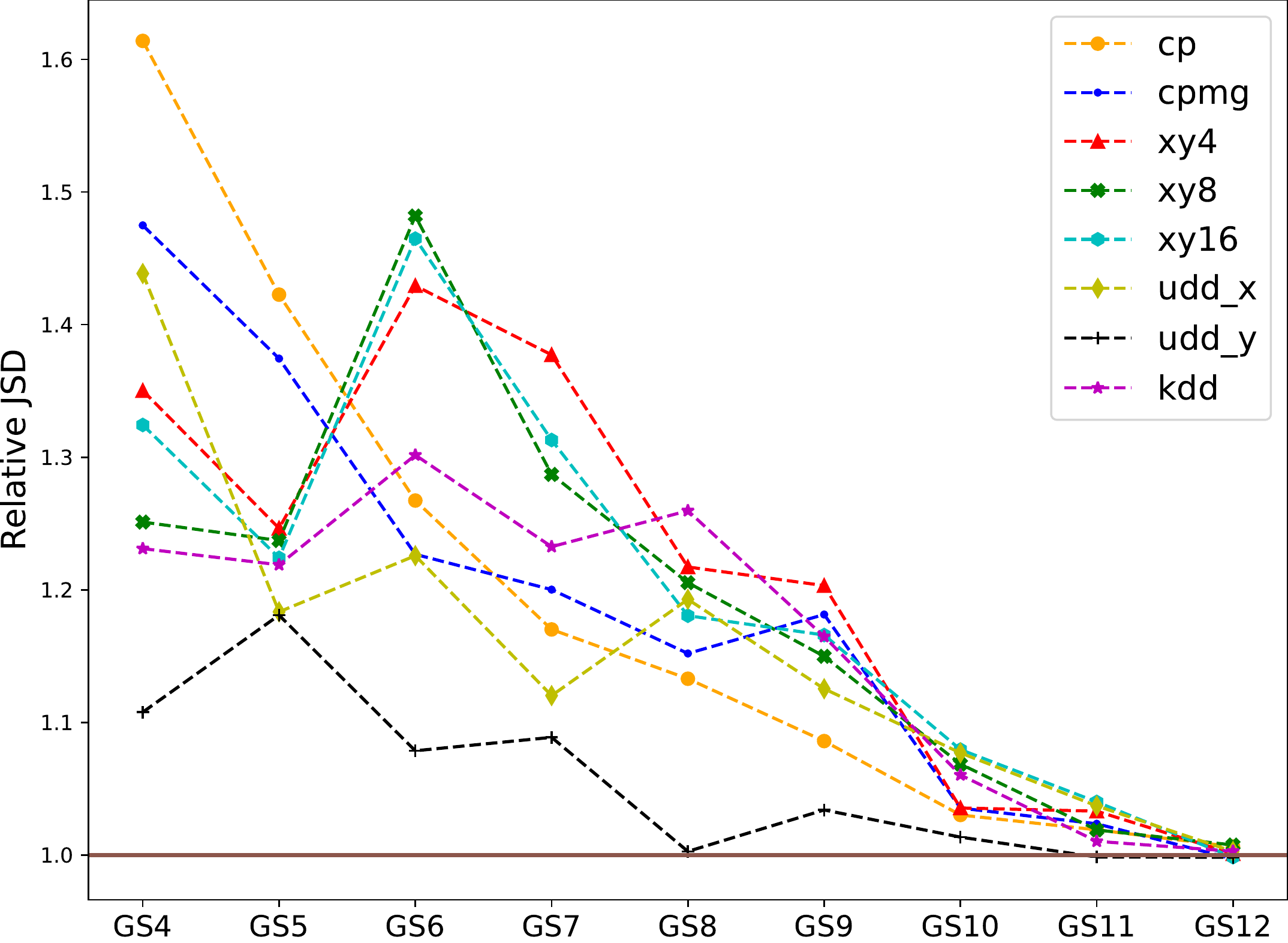}}

	\subfloat[ibmq\_toronto\label{gs_toronto}]{	\includegraphics[scale=0.28,width=0.4\textwidth]{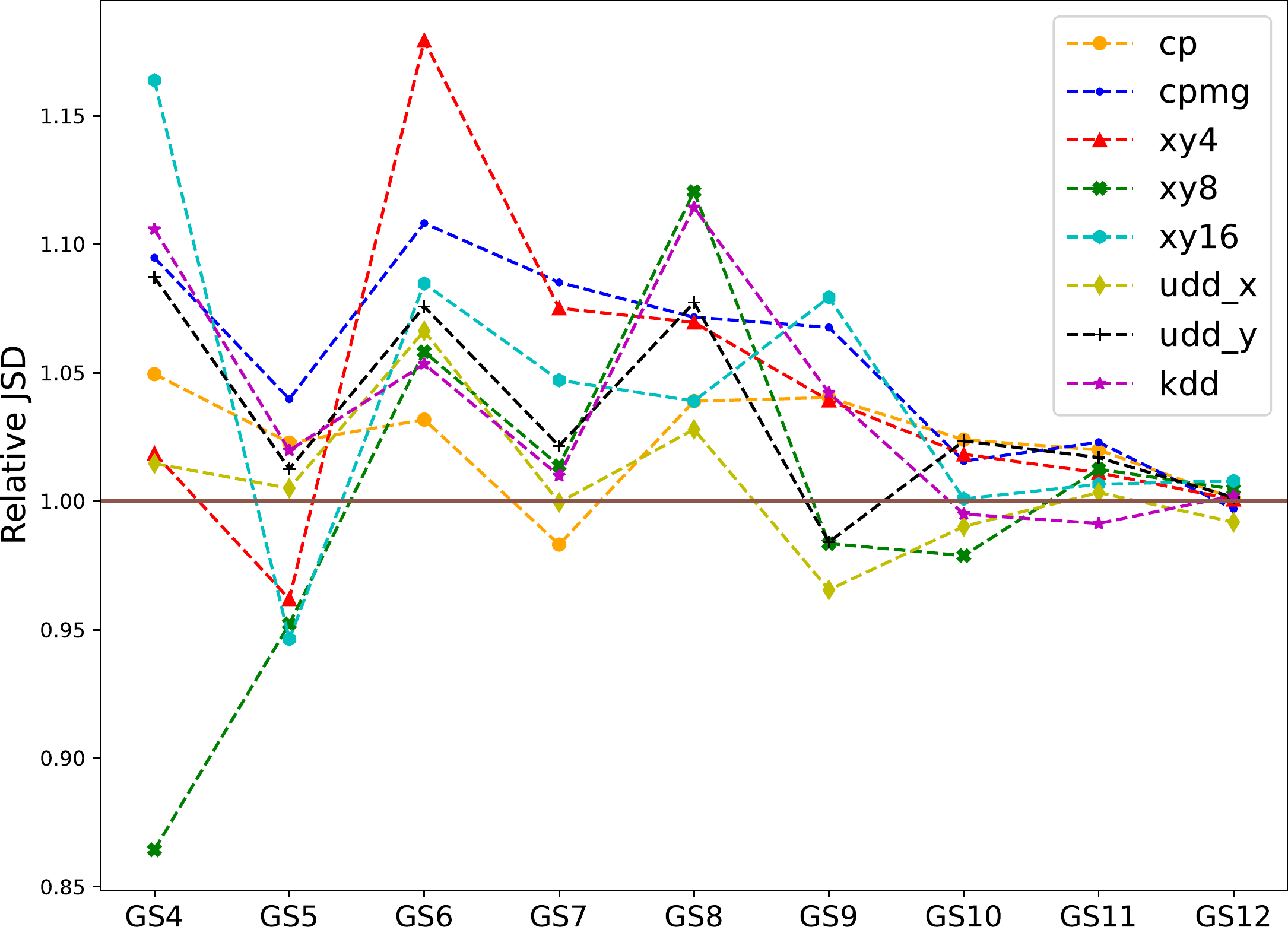}}
\hfil
	\subfloat[ibmq\_montreal\label{gs_montreal}]{	\includegraphics[scale=0.28,width=0.4\textwidth]{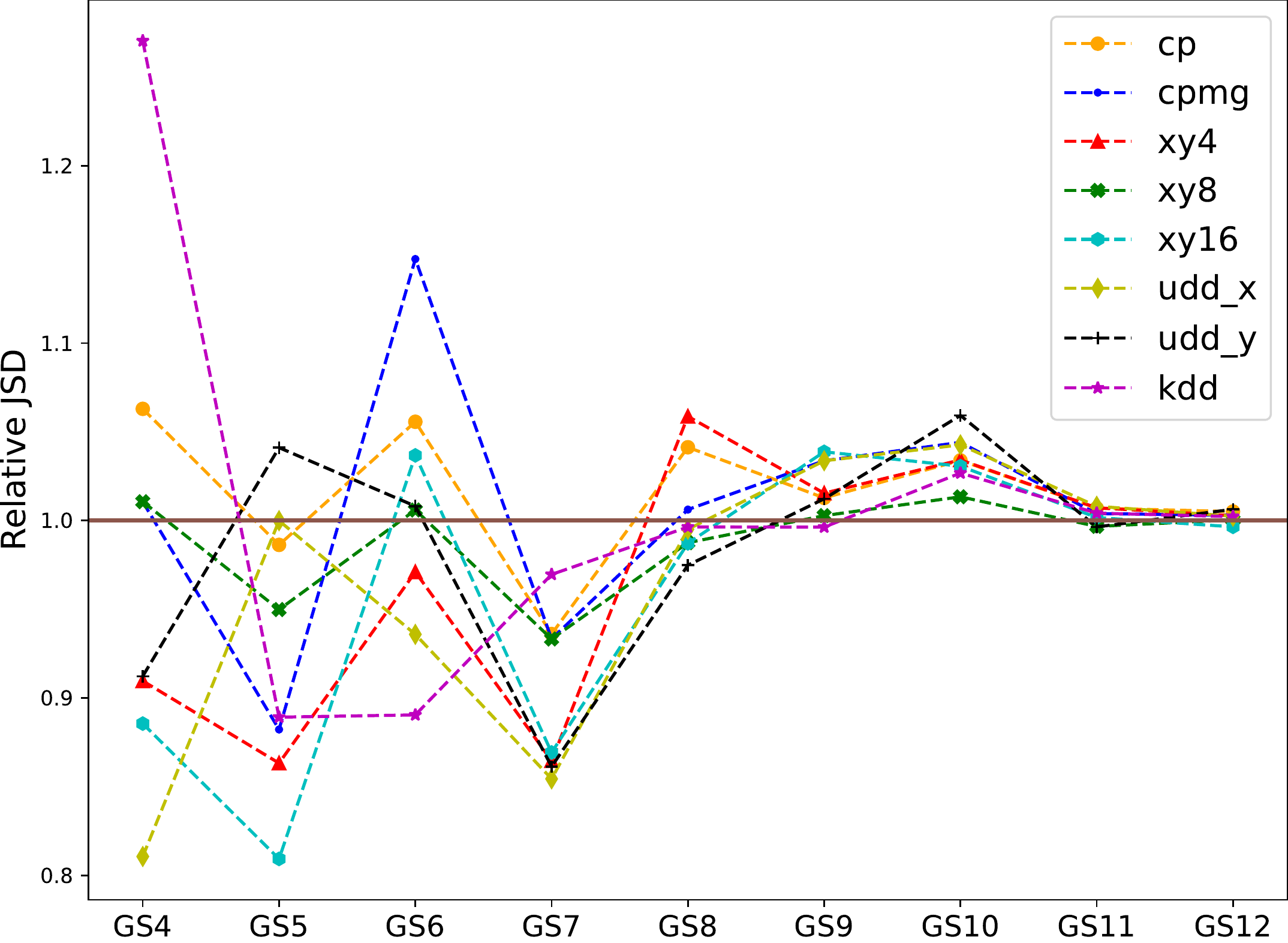}}

	\caption{Relative JSD results for Graph State circuits on different IBM quantum devices. Higher is better.}
	\label{fig:gss}
\end{figure*}

\begin{figure*}[h]
	\centering 
	\subfloat[ibmq\_jakarta\label{qaoa_jakarta}]{	\includegraphics[scale=0.28,width=0.4\textwidth]{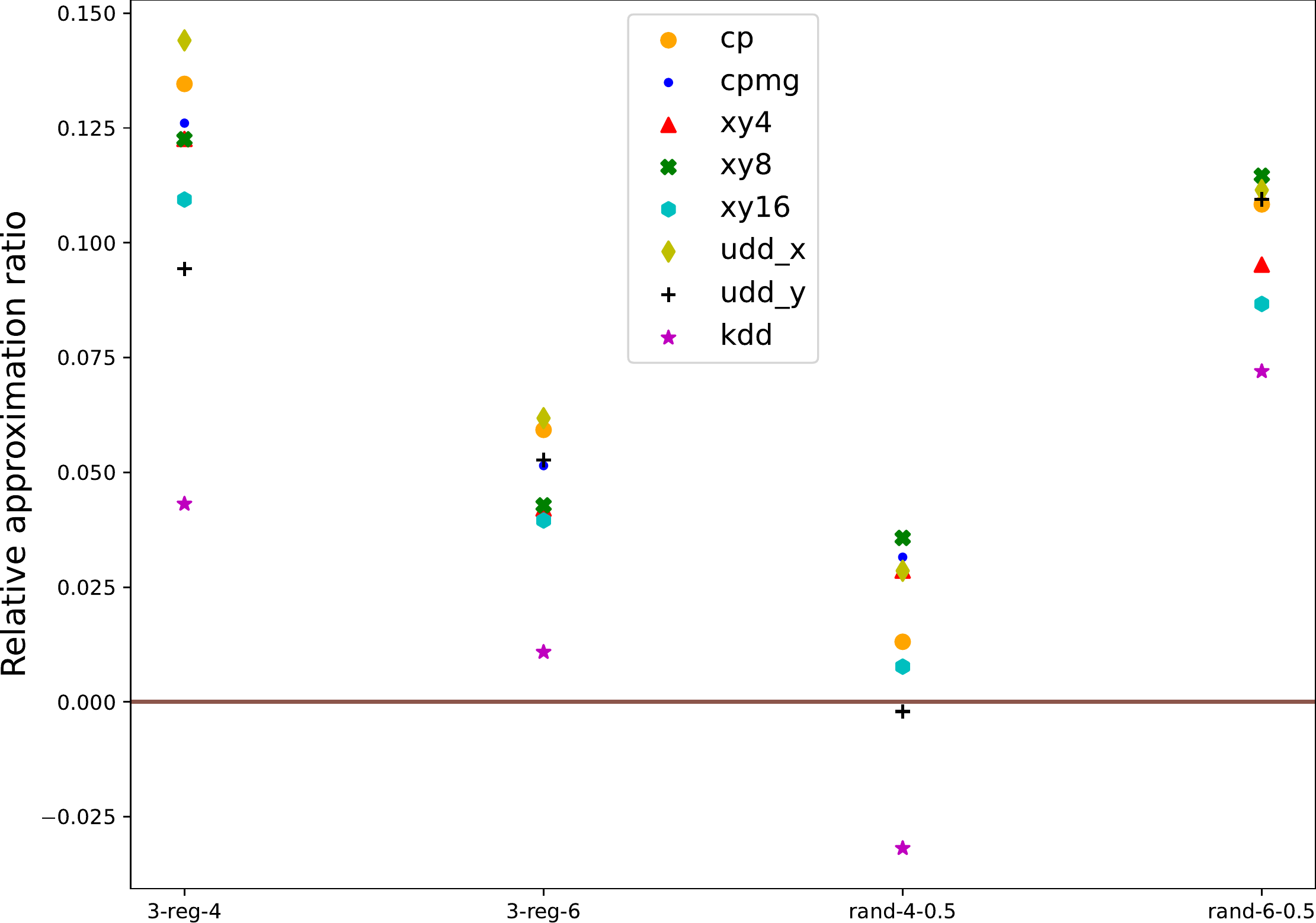}}
\hfil
	\subfloat[ibmq\_guadalupe\label{qaoa_guadalupe}]{\includegraphics[scale=0.28,width=0.4\textwidth]{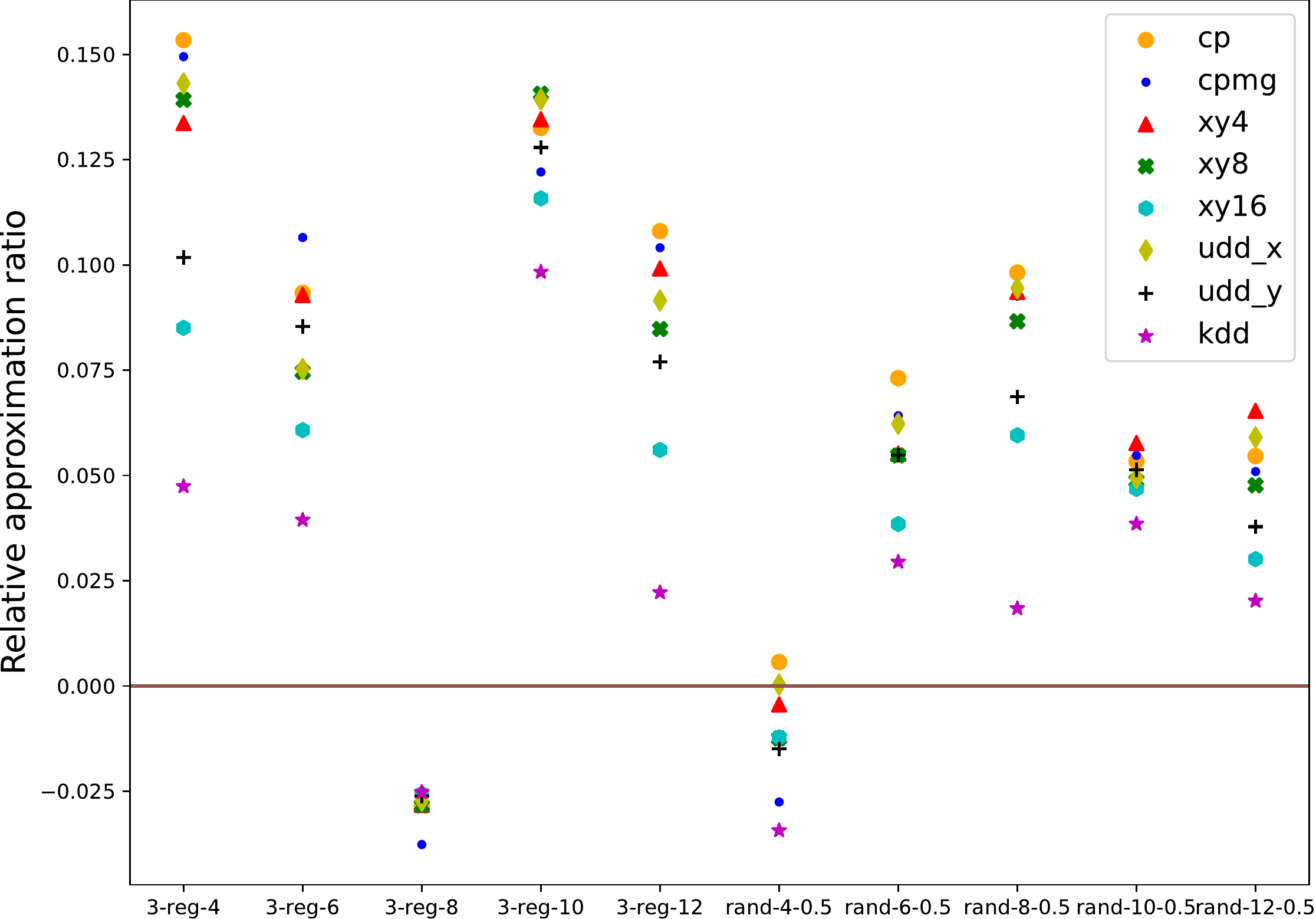}}

	\subfloat[ibmq\_toronto\label{qaoa_toronto}]{\includegraphics[scale=0.28,width=0.4\textwidth]{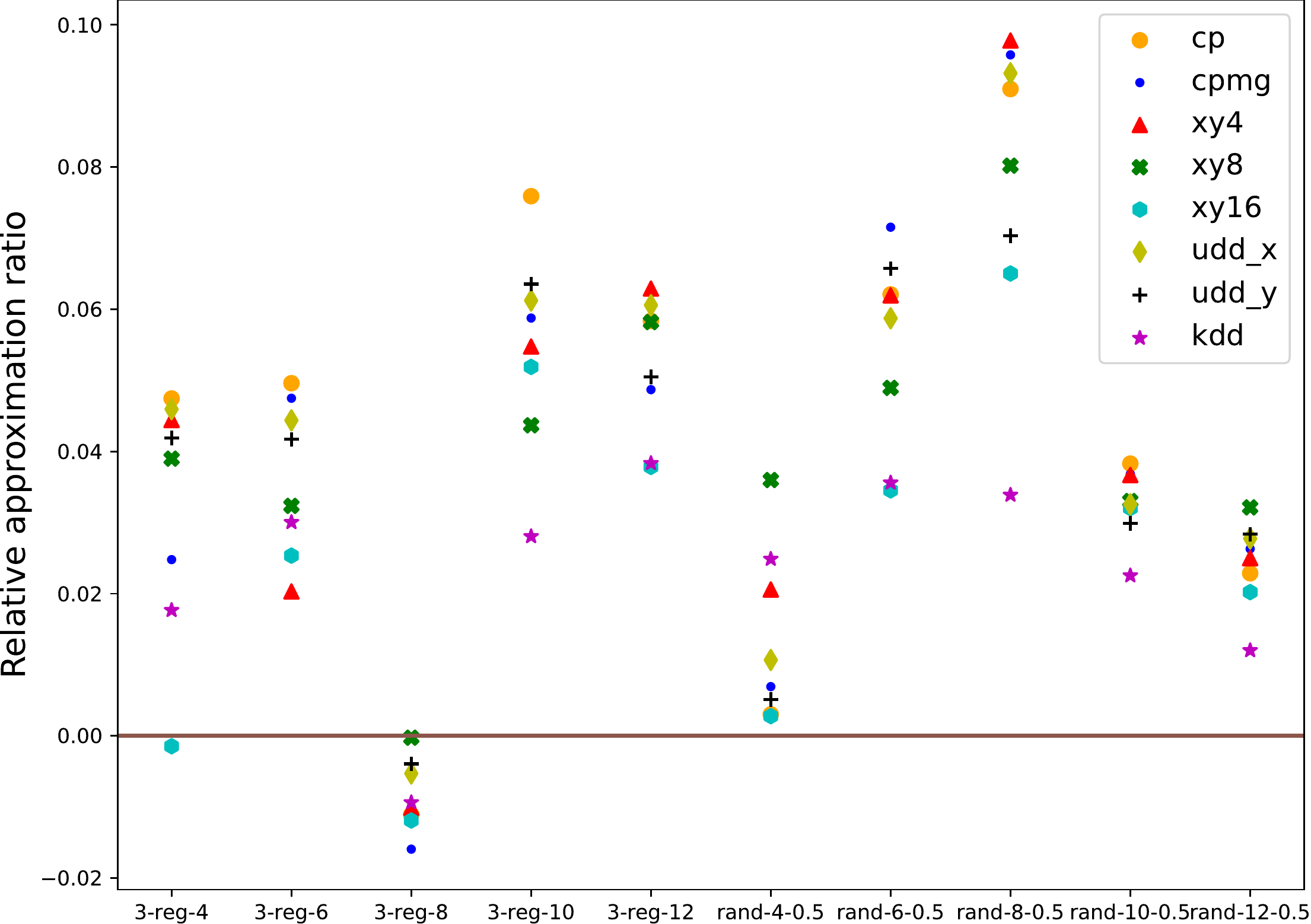}}
\hfil
	\subfloat[ibmq\_montreal\label{qaoa_montreal}]{\includegraphics[scale=0.28,width=0.4\textwidth]{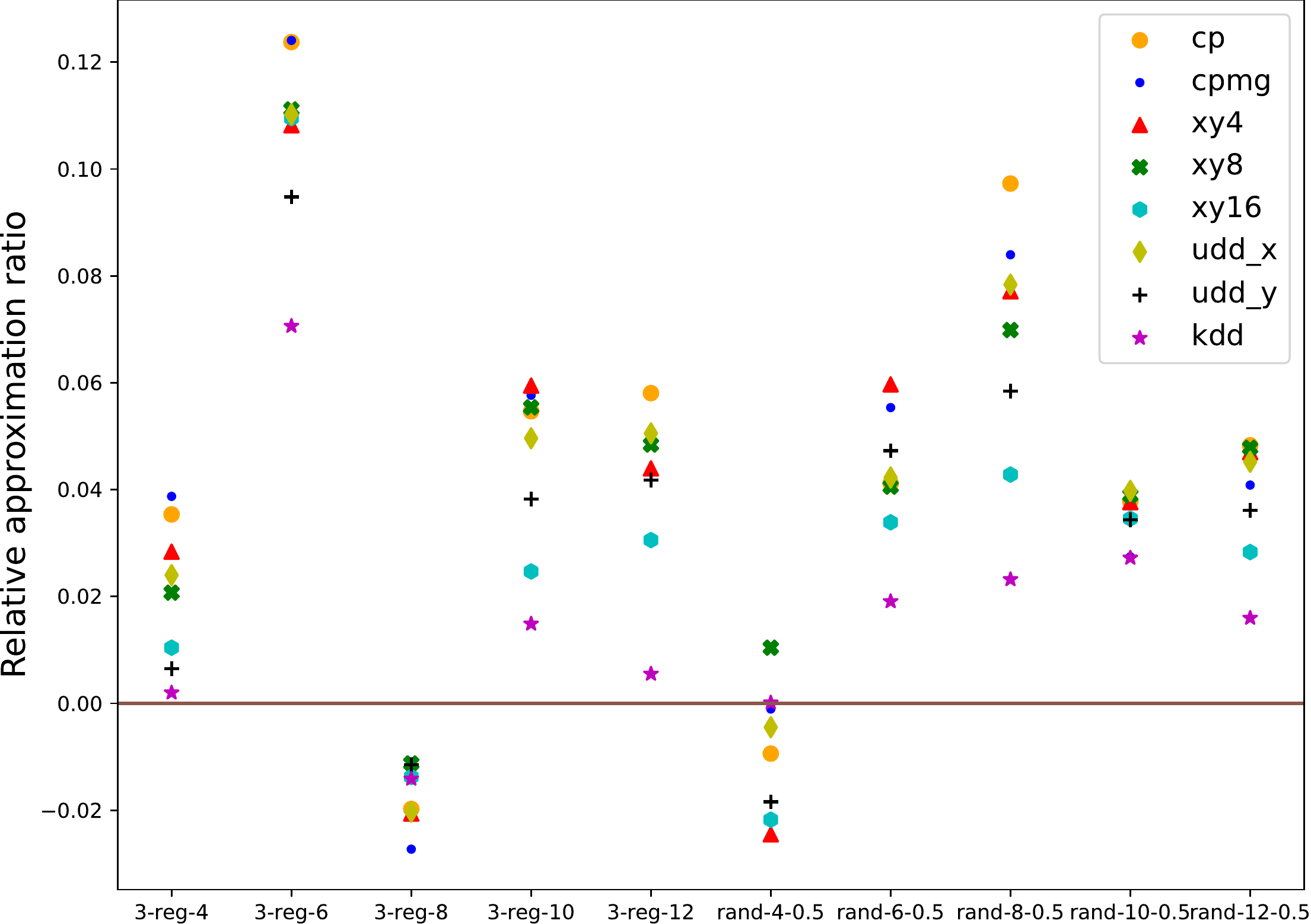}}

	\caption{Relative approximation ratio of applying DD sequences for QAOA circuit on different IBM quantum devices. Higher is better.}
	\label{fig:qaoa_dd}
\end{figure*}

\begin{figure*}[h]
	\centering 
	\subfloat[ibmq\_jakarta\label{qaoa_jakarta_pe}]{	\includegraphics[scale=0.28,width=0.4\textwidth]{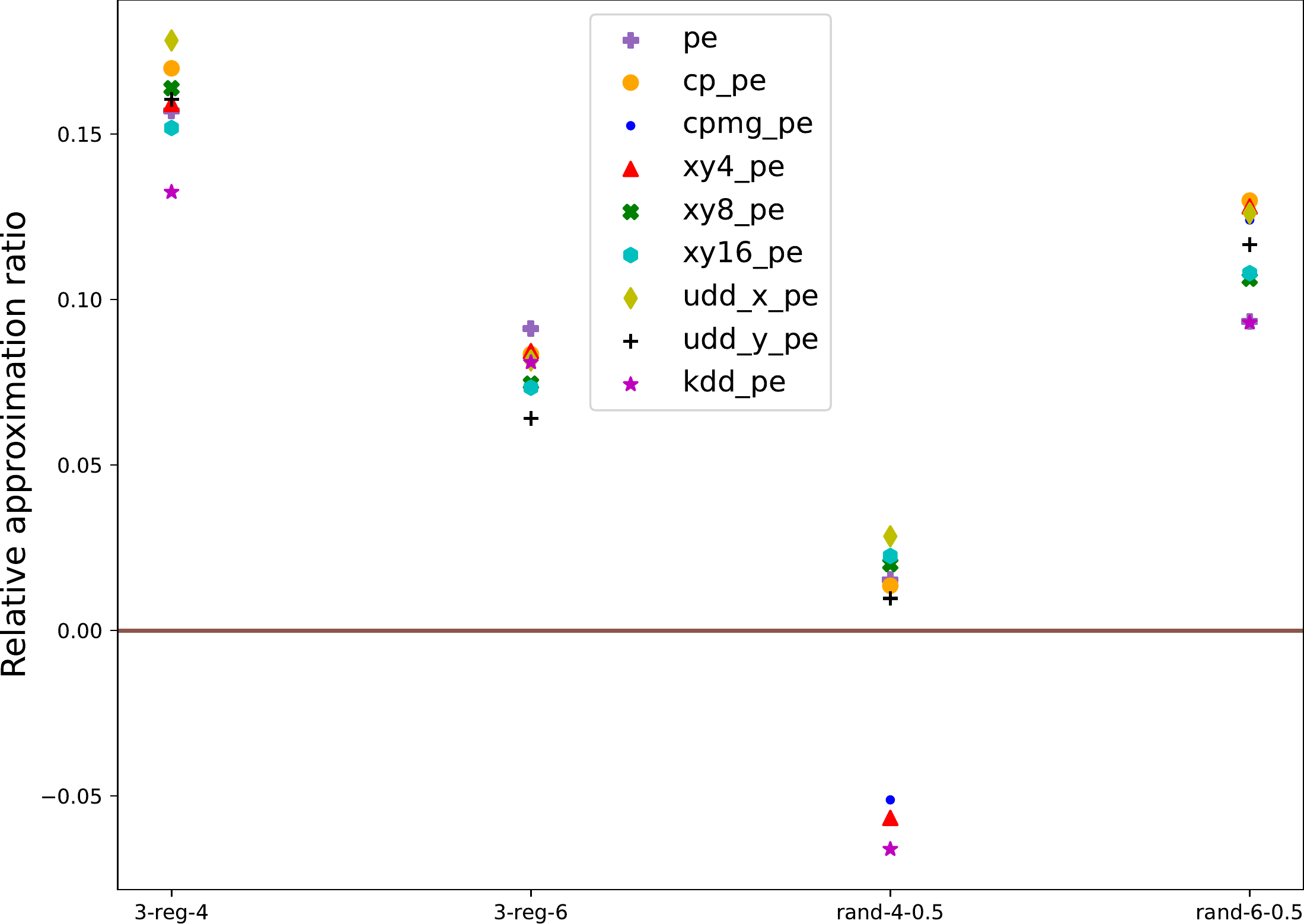}}
\hfil
	\subfloat[ibmq\_guadalupe\label{qapa_guadalupe_pe}]{\includegraphics[scale=0.28,width=0.4\textwidth]{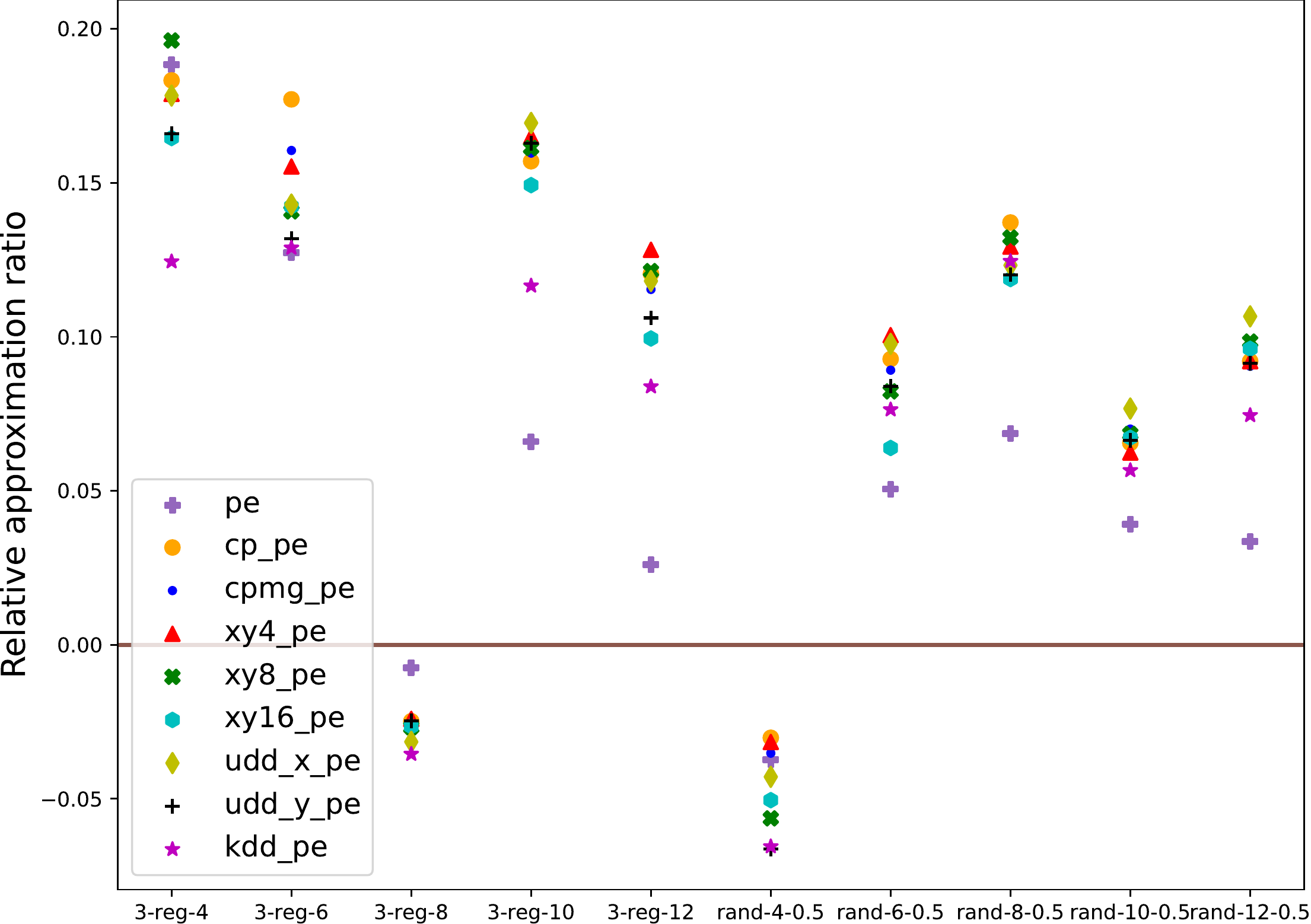}}
	
	\subfloat[ibmq\_toronto\label{qaoa_toronto_pe}]{\includegraphics[scale=0.28,width=0.4\textwidth]{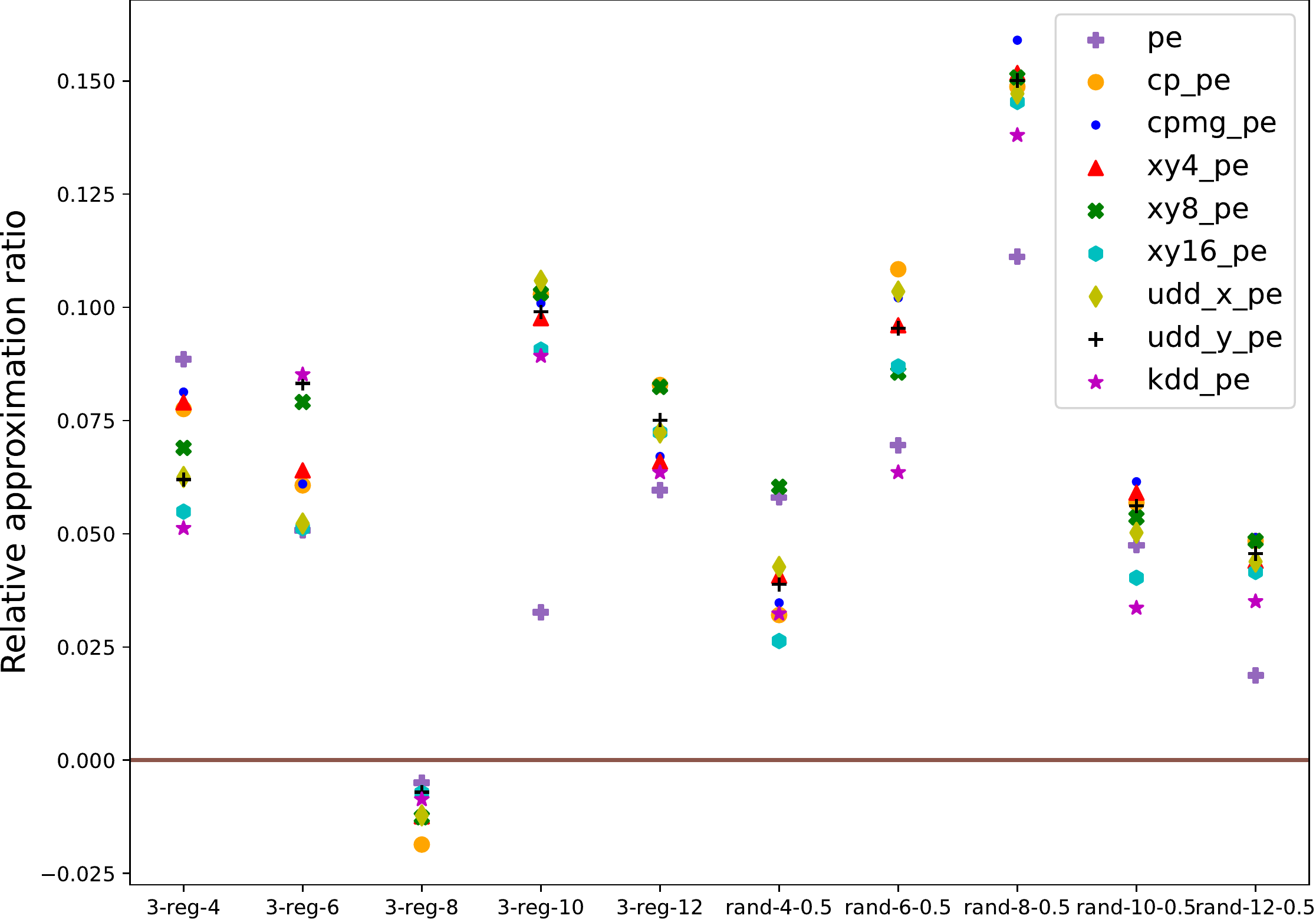}}
\hfil
	\subfloat[ibmq\_montreal\label{qaoa_montreal_pe}]{\includegraphics[scale=0.28,width=0.4\textwidth]{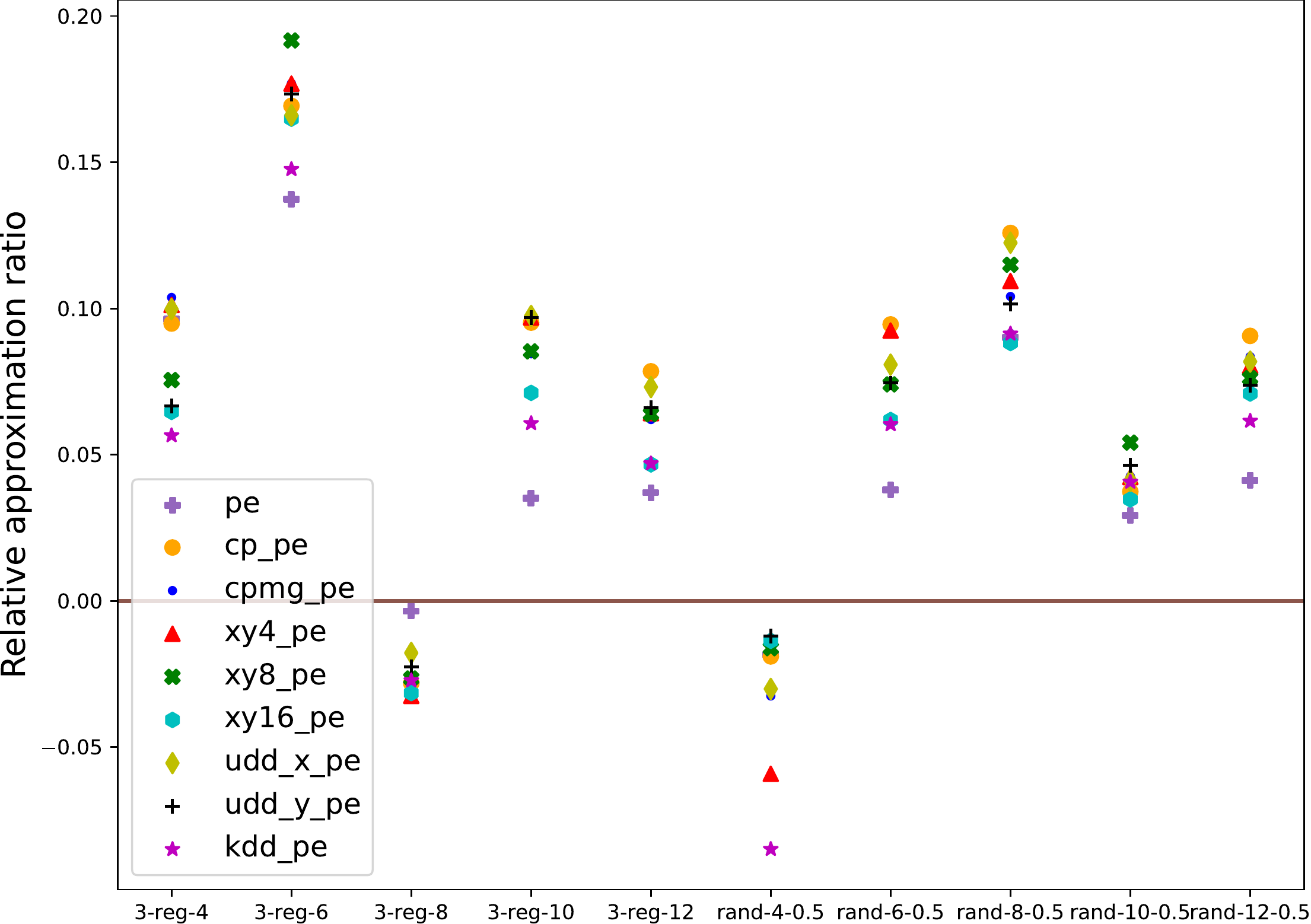}}

	\caption{Relative approximation ratio of applying DD sequences + pulse-efficient technique for QAOA circuit on different IBM quantum devices. Higher is better.}
	\label{fig:qaoa_dd_pe}
\end{figure*}

\begin{table}[!htp]
	\caption{\label{tab:ibmq}The information of different IBM Q devices.}
	
	\begin{tabular}{c c c c c}
		\toprule                              
		IBM Q & Jakarta & Guadalupe & Toronto & Montreal \cr
		\midrule
		$n$ & 7 & 16 & 27 & 27  \\
		\midrule
		$QV$ & 16 & 16 & 32 & 128   \\
		\bottomrule
	\end{tabular}
	
\end{table}

\section{Experimental results}
We perform the experiments on various IBM quantum devices with different qubit numbers and quantum volumes (QV) (see Table~\ref{tab:ibmq}). The size of the benchmarks varies depending on the quantum device. For the first experiment when evaluating the effects of applying DD sequences to different applications including BV, HS, QFT, and GS, the circuit size varies from 3 to 6 for IBM Q 7 Jakarta. Whereas for other devices, the circuit size changes from 3 to 12, since there are too many noises accumulating to obtain meaningful results for benchmarks with more than 12 qubits. When exploring the performance of combining DD and pulse-efficient optimization technique on QAOA for the second experiment, the degree of the two types of graphs (3-regular graph and random graph) ranges from 4 to 6 for IBM Q 7 Jakarta, and 4 to 12 for other devices.  The QAOA ansatz has one layer with parameters initialized randomly and optimized using \emph{COBYLA} optimizer on the simulator. We only execute the final ansatz with optimized parameters on the quantum hardware. All the benchmarks are compiled using Qiskit with the highest optimization level and executed 8192 times.


There are some limitations when inserting certain DD pulses to the idle time. If the DD sequences contain a large number of pulses, such as XY16 and KDD, it might not be possible to insert them to some small benchmarks whose idle time might not be long enough. Moreover, Hahn echo can only be applied if the inverse of the inserted single $X$ or $Y$ gate is able to be absorbed into the neighboring gates to ensure the equivalence of the quantum state. Therefore, we check if the DD pulses are actually inserted to the benchmark for each experiment, and we remove the circuit without any DD pulses inserted from the result.


In order to clearly show the impact of various DD sequences and pulse-efficient optimization method, we use the relative results for the three metrics. The original benchmark with no optimization method applied is marked as the baseline. PST is divided by the baseline and we use JSD results to divide the baseline, so that we can obtain the relative results when applying DD sequences. If the relative result is larger than one, it means that there is an improvement, and larger is better. Whereas for QAOA experiments, we use the difference between approximation ratio with DD sequence or pulse-efficient technique and the baseline as the relative result due to the possible negative value of the approximation ratio. If the difference is larger than zero, it indicates an enhancement and also larger is better. Each experiment has been repeated three times and results with similar trends were obtained. We show the average of the three experiments for all the results.


We first demonstrate applying DD sequences to various quantum applications. Hahn echo is not applicable for all the benchmarks. XY16 and KDD are too long to insert for certain small BV circuits (less than 4 qubits). All the DD sequences cannot be applied to HS algorithm, since the duration of the idle time is always equal to the duration of one single-qubit gate. Thus, the idle time is too short to fit any DD sequences. The relative PST results for BV circuits, the relative JSD results for QFT and Graph State circuits are shown in Fig.~\ref{fig:bvs}, Fig.~\ref{fig:qfts}, and Fig.~\ref{fig:gss}.

The quantitative analysis of the relative results is shown in Table~\ref{tab:3}. For BV algorithm (see Fig.~\ref{fig:bvs}), if the BV circuit involves a large number of qubits, the PST fidelities can be dropped dramatically such that all the PSTs are below 0.1 even with DD applied for error mitigation. Therefore, we only show the results whose PSTs are larger than 0.1. The relative PST results demonstrate that all the DD sequences are able to enhance the BV circuit fidelity, especially UDD\_X. On average, the fidelity is improved by 1.09x, 3.82x, 2.79x, and 1.68x compared with the baseline for IBM Q Jakarta, Guadalupe, Toronto, and Montreal respectively. Whereas for QFT, inserting DD sequence is more favorable for circuits with more than 5 qubits. Therefore, the relative JSD results on IBM Q 7 Jakarta are not encouraging. Overall, DD can be beneficial for QFT circuits but are not as stable as for BV circuits. The fidelity is improved by 1.14x, 1.18x, and 1.28x on average for IBM Q Guadalupe, Toronto, and Montreal. For Graph State circuit, the performance of DD is also not always steady and different across IBM quantum devices. DD is helpful for Graph State circuits on IBM Q Guadalupe but not for other devices, with an increase of fidelity by 1.17x on average. The variance of the results for different DD techniques becomes smaller for relatively large-scale QFT and Graph State circuits which might be due to the accumulating errors.



To conclude, DD can always improve the fidelity for BV algorithm to different extents across IBM quantum chips. But for other applications, the performance of DD remains uncertain and can be different on various quantum devices. Moreover, for certain benchmark like HS algorithm, the idle time is always short so that DD is not applicable regardless of the circuit size.

Second, we show the results of using QAOA to solve Max-Cut problem for different graphs by employing: (1) diverse DD sequences (see Fig.~\ref{fig:qaoa_dd}), (2) DD sequences + pulse-efficient optimization technique (see Fig.~\ref{fig:qaoa_dd_pe}). The quantitative analysis of the relative results is shown in Table~\ref{tab:4}. Most of the DD sequences can help enhance the approximate ratio for QAOA, especially CP sequence. The approximate ratio of all the graphs is improved by 6.8\%, 5.9\%, 3.7\%, and 3.7\% on average for IBM Q Jakarta, Guadalupe, Toronto, and Montreal respectively. There is no clear relationship between the level of improvement and the size of the benchmark. Comparing to the baseline circuit without applying any error mitigation technique, purely the pulse-efficient method is already able to raise the approximation value by 8.9\%, 5.5\%, 5.3\%, and 4.8\% on average for the four devices. Associating pulse-efficient technique with DD sequences can further improve the approximation ratio and the negative impact induced by DD can mostly be canceled. The combination of DD and pulse-efficient technique can improve the approximation ratio by 26.5\%, 49.2\%, 81.1\%, 75.7\% compared with only applying DD for the four IBM devices.

%

\begin {table*}[ht]
\caption{\label{tab:3}The relative results of applying various DD sequences to different quantum applications on four IBM Q hardware.}
\begin{center}
	\resizebox{\linewidth}{!}{%
		\begin{threeparttable}
	\begin{tabular}{|l|l|l|l|l|l|l|l|l|l|l|l|l|}
		\hline 
		\multirow{3}{*}{Benchmarks} & \multicolumn{12}{|c|}{IBM Q Hardware} \\
		\cline{2-13}
		& \multicolumn{3}{|c|}{Jakarta} & \multicolumn{3}{|c|}{Guadalupe} & \multicolumn{3}{|c|}{Toronto} & \multicolumn{3}{|c|}{Montreal} \\
		\cline{2-13}
		& Max & Min & Avg & Max & Min & Avg & Max & Min & Avg & Max & Min & Avg \\
		\hline
		BV & 1.1 (UDD\_X) & 1.05 (KDD) & 1.09 & 4.53 (UDD\_X) & 3.16 (CPMG) & 3.82 & 3.18 (UDD\_X) & 2.44(XY16) & 2.79 & 1.77(UDD\_X) & 1.55 (KDD) & 1.68 \\
		\hline
		QFT & 1.1 (XY16) & 0.89 (UDD\_Y) & 1.01 & 1.17 (CPMG) & 1.06 (UDD\_X) & 1.14 & 1.23 (KDD) & 1.14 (UDD\_X) & 1.18 & 1.33 (XY4) & 1.22 (UDD\_X) & 1.28 \\
		\hline
		GS & 1 (CP) & 0.9 (UDD\_Y) & 0.95 & 1.21 (XY4) & 1.06 (UDD\_Y) & 1.17 & 1.06 (CPMG) & 1 (XY8) & 1.03 & 1.02 (CP) & 0.96 (XY16) & 0.99\\
		\hline 
		HS & \multicolumn{12}{|c|}{DD is not applicable}\\
		\hline
	\end{tabular}
\begin{tablenotes}
	\item If the relative results > 1, it means that the DD sequence has a positive impact on the fidelity . Otherwise, a negative impact is introduced by DD. 
\end{tablenotes}

\end{threeparttable}
}
\end{center}

\end{table*}

One interesting thing is that, even though KDD is a robust DD sequence that is designed to be robust against pulse imperfection and has shown improvements on Rigetti device, its performance is not as good as expected on IBM quantum devices. For all the quantum applications tested including QAOA circuit, there is no growth of circuit fidelity using KDD compared with other DD strategies and actually it decreases the circuit fidelity for most of the cases.

\begin {table*}[ht]
\caption{\label{tab:4}The relative results of applying different methods, including only DD, only pulse-efficient method (PE), and DD+PE to QAOA circuits on four IBM Q hardware.}
\begin{center}
	\resizebox{\linewidth}{!}{%
		\begin{threeparttable}
			\begin{tabular}{|l|l|l|l|l|l|l|l|l|l|l|l|l|}
				\hline 
				\multirow{3}{*}{Methods} & \multicolumn{12}{|c|}{IBM Q Hardware} \\
				\cline{2-13}
				& \multicolumn{3}{|c|}{Jakarta} & \multicolumn{3}{|c|}{Guadalupe} & \multicolumn{3}{|c|}{Toronto} & \multicolumn{3}{|c|}{Montreal} \\
				\cline{2-13}
				& Max (\%)& Min (\%)& Avg (\%)& Max (\%) & Min (\%)& Avg (\%)& Max (\%)& Min (\%)& Avg (\%)& Max (\%)& Min (\%)& Avg (\%) \\
				\hline
				Only DD & 8.6 (UDD\_X) & 2.4 (KDD) & 6.8 & 7.4 (CP) & 2.5 (KDD) & 5.9 & 4.4 (CP) & 2.3 (KDD) & 3.7 & 4.7 (CP) & 1.6 (KDD) & 3.7 \\
				\hline
				Only PE & \multicolumn{3}{|c|}{8.9} & \multicolumn{3}{|c|}{5.5} &\multicolumn{3}{|c|}{5.3} & \multicolumn{3}{|c|}{4.8} \\
				\hline
				DD+PE & 10.4 (UDD\_X) & 6 (KDD) & 8.6 & 9.7 (CP) & 6.8 (KDD) & 8.8 & 7.2 (XY8) & 5.8 (KDD) & 6.7 & 7.4 (CP) & 4.5 (KDD) & 6.5 \\
				\hline
			\end{tabular}
			\begin{tablenotes}
				\item If the relative results > 0, it means an improvement on fidelity. Otherwise, a negative impact is introduced on fidelity.
			\end{tablenotes}
			
		\end{threeparttable}
	}
\end{center}

\end{table*}
\section{Discussion}
\subsection{Design guideline}
It is important to explore different noise mitigation and pulse-level optimization methods, including DD and pulse-efficient optimization technique to better build the quantum circuits.
Here, we list several guidelines that can help the  community utilize these methods to design circuits with higher fidelity.
\begin{itemize}
	\item DD cannot always improve the circuit fidelity and it is highly application-dependent. It is recommended to use DD on specific quantum algorithms, including Bernstein-Vazirani algorithm, and QAOA.  
	
	\item When applying DD sequences, it is recommended to check the transpiled circuit to verify if DD is suitable to insert.
	
	\item The robust sequence KDD does not work well on IBM quantum devices for most of the quantum algorithms tested. 
	
	\item The pulse-efficient optimization technique is favorable for QAOA circuits and combined with DD, it can improve the approximation ratio.

\end{itemize}
\subsection{Future work}
As the experiments have demonstrated some discouraging effects when inserting DD sequences for certain quantum applications, exploring the hardware physics behind them becomes important. Application-dependent adaptive dynamical decoupling~\cite{das2021adapt} and variational algorithm-focused VAQEM~\cite{ravi2021vaqem} have been proposed to provide the most beneficial DD sequences but with large overhead and limited analysis of different DD strategies. A smarter approach for designing appropriate application-oriented DD sequences considering hardware physics is left to future works. Recently, DD has been proven to be able to suppress ZZ-crosstalk for fixed frequency transmon superconducting device~\cite{tripathi2021suppression}. It is interesting to further investigate the performance of DD-based crosstalk suppression at application-level, so that it can contribute to the quantum parallel executions~\cite{niu2021enabling,niu2021parallel}. Finally, the pulse-level optimization experiments can be delved to other quantum applications to provide more guidelines, such as VQE~\cite{kandala2017hardware}, quantum simulation~\cite{georgescu2014quantum}, etc.

\section{Conclusion}
Today's quantum hardware is prone to noise in the NISQ era. Therefore, circuit optimization and error mitigation approaches are required to increase the output fidelity. In this paper, we focus on two pulse-level circuit optimization methods: dynamical decoupling and pulse-efficient optimization technique. First, we implement various DD strategies on IBM quantum devices including non-universal, universal, and robust ones. Second, we apply these DD sequences to several well-known quantum applications, such as QFT and Bernstein-Vazirani algorithm, to evaluate comprehensively the impact of diverse DD techniques on IBM quantum devices with different characteristics. We also merge DD with pulse-efficient transpilation method and investigate them on QAOA circuits to solve Max-Cut problem. Based on the experimental results, we found that DD techniques always show positive impact for some benchmarks, while for others, DD demonstrate some discouraging effects. As there is no overhead-free application-oriented DD approach, we provide a list of design guidelines for users to better understand these pulse-level optimization methods and figure out how to improve the circuit design for various quantum applications.

\section*{Acknowledgment}
This work is funded by the QuantUM Initiative of the Region Occitanie, University of Montpellier and IBM Montpellier. The authors are very grateful to Adrien Suau for the helpful suggestions and feedback on an early version of this manuscript.

\section*{Supplementary Information}\label{supplementary}

All the scripts to reproduce the experimental results are publicly available in \url{https://github.com/peachnuts/DD_PE}.

\bibliography{apssamp}

\end{document}